\documentclass[8pt]{article}

\usepackage{sectsty}
\usepackage[pdftex]{graphicx}
\usepackage{bm}
\usepackage{color}
\usepackage{textcase, url}
\usepackage[normalem]{ulem}
\usepackage[colorlinks=true, linkcolor=black, urlcolor=black, citecolor=black]{hyperref}
\usepackage{amsmath,amssymb}
\usepackage{mathtools}
\usepackage{lineno}
\usepackage{tabularx}
\usepackage{booktabs}
\usepackage[a4paper,bottom=20mm,right=20mm,left=20mm,top=20mm]{geometry}
\usepackage[super,comma, sort&compress]{natbib}
\usepackage{float}
\usepackage{indentfirst}
\usepackage[separate-uncertainty = true,multi-part-units=single]{siunitx}
\usepackage[margin=0mm]{caption}
\allsectionsfont{\normalfont\sffamily\bfseries}

\usepackage{amsmath}
\usepackage{amssymb}
\usepackage{amsthm}
\usepackage{amsfonts}
\usepackage{braket}

\usepackage{amsthm,amsmath}
\usepackage[utf8]{inputenc} %unicode support
\usepackage{siunitx}
\newcommand{\SIadj}[2]{\SI[number-unit-product={\text{-}}]{#1}{#2}}
\usepackage[pdftex]{graphicx}

\definecolor{darkblue}{rgb}{0,0,0.5}
\definecolor{darkgreen}{rgb}{0,0.5,0}
\definecolor{darkred}{rgb}{.7,0,0}
\definecolor{purple}{rgb}{0.5,0,0.6}
\definecolor{orange}{rgb}{1,0.5,0}
\definecolor{grey}{rgb}{.6,.6,.6}
\definecolor{lightpink}{rgb}{1,0.7,0.75}
\definecolor{pink}{rgb}{1,0.4,0.58}
\definecolor{deeppink}{rgb}{1,0.08,0.58}

\newcommand{\he}[1]{{\color{black}{#1}}}

\begin{document}
	
\begin{center}
	%%% T I T L E %%%
	% 	\textsf{\textbf{\LARGE Electron shuttling by a solitary acoustic chirped pulse}}\\\vspace{2mm}
	% 	\textsf{\textbf{\LARGE Solitary acoustic pulse synthesis for single-electron transport}}\\\vspace{2mm}
	\textsf{\textbf{\LARGE Semiconductor-based electron flying qubits: Review on recent progress accelerated by numerical modelling}}\\\vspace{2mm}
	% 	\textsf{\textbf{\LARGE Solitary acoustic pulse via chirp synthesis for single-electron transport}}\\\vspace{2mm}
	% 	\comment{HE: On my opinion, the title is now quite hard to read (try it). I think something more concise would be better. For instance: "Electron shuttling by (solitary) acoustic chirp."}\\
	{\small
		Hermann Edlbauer$^{1}$,
		Junliang Wang$^{1}$,
		Thierry Crozes$^{1}$,
		Pierre Perrier$^{1}$,
		Seddik Ouacel$^{1}$,
		Cl{\'e}ment Geffroy$^{1}$,
		Giorgos Georgiou$^{1,2,3}$,
		Eleni Chatzikyriakou$^{4}$,
		Antonio Lacerda-Santos$^{4}$,
		Xavier Waintal$^{4}$,
		D. Christian Glattli$^{5}$,
		Preden Roulleau$^{5}$,
		Jayshankar Nath$^{5}$,
		Masaya Kataoka$^{6}$,
		Janine Splettstoesser$^{7}$,
		Matteo Acciai$^{7}$,
		Maria Cecilia da Silva Figueira$^{8}$,
		Kemal {\"O}ztas$^{8}$,
		Alex Trellakis$^{8}$,
		Thomas Grange$^{9}$,
		Oleg M. Yevtushenko$^{8}$,
		Stefan Birner$^{8*}$  \&\\
		Christopher B{\"a}uerle$^{1*}$
	}
\end{center}
\vspace{-2mm}
\def\einr{2mm}
\def\spazi{-1.7mm}
{\small
	\begin{nolinenumbers}
		\-\hspace{\einr}$^1$ Univ. Grenoble Alpes, CNRS, Grenoble INP, Institut N\'eel, 38000 Grenoble, France\vspace{\spazi}\\
		\-\hspace{\einr}$^2$ Univ. Grenoble Alpes, CEA, Leti, 38000 Grenoble, France\vspace{\spazi}\\
		\-\hspace{\einr}$^3$ James Watt School of Engineering, Electronics and Nanoscale Engineering, University of Glasgow, University Avenue,\vspace{-2mm}\\ \-\hspace{3.5mm} Glasgow G12 8QQ, United Kingdom\vspace{\spazi}\\
		\-\hspace{\einr}$^4$ PHELIQS, Université Grenoble Alpes, CEA, Grenoble INP, IRIG, Grenoble 38000, France\vspace{\spazi}\\
		\-\hspace{\einr}$^5$ SPEC, Universit{\'e} Paris-Saclay, CEA, CNRS, Gif sur Yvette 91191, France\vspace{\spazi}\\
		\-\hspace{\einr}$^6$ National Physical Laboratory, Hampton Road, Teddington, Middlesex TW11 0LW, United Kingdom\vspace{\spazi}\\
		\-\hspace{\einr}$^7$ Department of Microtechnology and Nanoscience (MC2), Chalmers University of Technology, Göteborg, S-412 96,\vspace{-2mm}\\ \-\hspace{3.5mm} Sweden\vspace{\spazi}\\
		\-\hspace{\einr}$^8$ nextnano GmbH,
		Konrad-Zuse-Platz 8, 81829
		M{\"u}nchen, Germany\vspace{\spazi}\\
		\-\hspace{\einr}$^9$ nextnano Lab,
12 chemin des prunelles, 38700
Corenc, France\vspace{\spazi}\\
		%\-\hspace{\einr}$^\dagger$ these authors contributed equally to this work.
		\-\hspace{\einr}$^*$ corresponding author:
		\href{mailto:stefan.birner@nextnano.com;  christopher.bauerle@neel.cnrs.fr}{stefan.birner@nextnano.com; christopher.bauerle@neel.cnrs.fr}
		\\
		\\
	\end{nolinenumbers}
}	

\section*{Abstract}

The progress of charge manipulation in semiconductor-based nanoscale devices
opened up a novel route to realise a flying qubit with a single electron.
In the present review, we introduce the concept of these electron flying qubits, discuss their most promising realisations and show how numerical simulations are applicable to accelerate experimental development cycles.
Addressing the technological challenges of flying qubits that are currently faced by academia and quantum enterprises, we underline the relevance of interdisciplinary cooperation to move emerging quantum industry forward.
The review consists of two main sections:

\textbf{Pathways towards the electron flying qubit:}
We address three routes of single-electron transport in GaAs-based devices focusing on surface acoustic waves, hot-electron emission from quantum dot pumps and Levitons.
For each approach, we discuss latest experimental results and point out how numerical simulations facilitate engineering the electron flying qubit.

\textbf{Numerical modelling of quantum devices:}
We review the full stack of numerical simulations needed for fabrication of the flying qubits.
Choosing appropriate models, examples of basic quantum mechanical simulations are explained in detail.
We discuss applications of open-source (KWANT) and the commercial (nextnano) platforms for modelling the flying qubits.
The discussion points out the large relevance of software tools to design quantum devices tailored for efficient operation.

\pagebreak

\section{Introduction}

Flying qubits are originally intended to serve as a communication link within a quantum computer \cite{divincenzo_2000} \he{and represent a vital part of global road-maps towards secure data transmission -- the so-called quantum internet \cite{Wehner_2018}.}
Recently, in-flight manipulations of photon-number states (so-called Fock states) have shown that the flying qubit architecture can also be applied as a stand-alone quantum processing unit \cite{peruzzo_2014}.
Owing to this progress on photonic quantum-computation approaches \cite{knill_2001,kok_2007}, flying qubits are typically associated with photons \cite{lodahl_review_2018}.
\he{Employing so-called “time multiplexing” architectures, photonic quantum computing can in principle scale up to millions of qubits.}
% \cb{With the addition of so-called “time multiplexing” architectures, photonic quantum computing can in principle scale up to millions of qubits.}% H.E.: Good point, but not sure if it would fit in better a bit later?
The strongly probabilistic nature of photonic two-qubit gates renders the far-reaching photonic coherence however a double-edged sword making the realisation of photonic quantum computing challenging.
Though, a flying-qubit architecture can also be built on the basis of other quantum systems such as the electron \cite{bauerle_2018}.
The charge of an electron causes Coulomb interaction with its electro-magnetic environment, \he{which}
% what \SB{Comment: 'which' instead of 'what'?}
exposes its quantum properties to decoherence but enables well-controlled single-particle manipulations and multi-qubit coupling.

% \OYe{"granular" -> "discreet"? "grain" -> quantum?}
\he{Historically, flying qubits stem from the research field of quantum optics which addresses -- in opposition to wave optics --  the granular nature of light.}
% Quantum optics addresses the granular nature of light as opposed to wave optics.
The need to control the ultimate grain of light -- a single photon -- \he{led}
% \OYe{leads?} # H.E.: It is simple past tense here ... so "led" is the correct conjugation.
to the advent of the first single-photon source in 1974 \cite{clauser_1974} and over several decades, various promising approaches have been developed \cite{eisaman_2011}.
\he{
	A prominent example are deterministic single-photon sources based on quantum dots (QD) \cite{michler_2000, senellart_2017b}.
}
% A prominent example are single-photon sources based on quantum dots (QD) \cite{michler_2000, senellart_2017b} \cb{as truly deterministic
% single-photon sources.}
Since these photon emitters allow for efficient coupling to a
\he{nanophotonic}
% nanophonotic
cavity \cite{lodahl_2014,lodahl_2020} and provide a high degree of indistinguishability \cite{yamamoto_2002, he_2013} and brightness \cite{munch_2013, somachi_2016,Tomm_2021},
%SB: I inserted a comma here.
they turned out as highly suitable sources for quantum communication \cite{lodahl_review_2018}.
% warburton_2021
%Tomm, N., Javadi, A., Antoniadis, N.O. et al. A bright and fast source of coherent single photons. Nat. Nanotechnol. 16, 399–403 (2021).
%
% \cbc{Connection to lithographically defined quantum dots which are used for single electron sources is done in section 2}
\he{
	Yet, the extraction efficiencies of quantum dot based single-photon sources are not at the level to perform photonic quantum computing.
	Nevertheless, the demand for photonic components such as on-demand single-photon sources, quantum photonic processors and single-photon detectors fostered the emergence of start-ups such as \textit{Quandela}\footnote{https://quandela.com},  \textit{QuiX}\footnote{https://www.quix.nl}, or \textit{SingleQuantum}\footnote{https://singlequantum.com}, to name a few.
	Photonic quantum computing is currently pursued by the start-up \textit{PsiQuantum}\footnote{https://psiquantum.com} which uses heralded photons.
	Here parametric down conversion is used to create from a non-deterministic single photon source a pair of photons serving as signal and idler.
	Measuring the idler photon with an additional single-photon detector ensures that indeed a single photon (signal) has been generated -- however, at the expense of additional hardware.
	% -- a single-photon detector. % H.E.: Maybe it's not necessary to go in that detail here ...
}

In order to implement photonic quantum computation, two-qubit gates are a necessity.
As photons do not interact, operating two-qubit gates on physical qubits is very difficult.
It is possible in theory to use non-linear effects such as Kerr effects, but these effects are so small that it is not possible in practice, at least in known media~\cite{kok_2007}.
Another route is to use photon detectors and the associated projective measurement as a way to entangle photons.
While photon detectors can indeed produce entanglement they do so in a probabilistic way, i.e. depending on the result of the measurement, the remaining state will be entangled or not.
% \he{HE: I don't understand this sentence.}
\he{However,} such probabilistic gates cannot be used on a large quantum computer as the probability for the correct circuit to be applied will decay exponentially with the number of two-qubit gates.
Protocols have been proposed (such as the Knill--Laflamme--Milburn protocol) \cite{knill_2001} trying to mitigate this issue and to enhance the probability for the correct gate to be applied.
This approach comes however along with significant cost since many ($n \gg 1$) ancilla photonic qubits are required to guarantee that the correct gate is applied (with an error probability of the order of $1/n$) \cite{kok_2007}.
From this point of view, the development of non-photonic approaches represents a promising and competitive avenue.

\he{
	For sake of completeness, let us also point out a different approach to photonic quantum computing based on so-called \textit{squeezed states}.
	Here the qubits are generated as a superposition of multiple photons in a light pulse.
	This approach is pursued by the start-up \textit{Xanadu}\footnote{https://www.xanadu.ai}.
	It has conceptual similarity to the Leviton single-electron transport which will be presented below in section 2.3.
}
% \cb{
% For sake of completeness, let us also point out that other approaches to photonic quantum computing exists where the photonic qubit consists of so-called \textit{squeezed states} generating a superpositions of multiple photons in a light pulse.
% This approach is pursued by the start-up \textit{Xanadu}\footnote{https://www.xanadu.ai}.
% As we will, see this is very similar to the Leviton single-electron source which will be presented} \oy{below} \cb{in section 2.3. % H.E.: From our explanation, the reader will only have a rough idea about heralded photons ... so we should avoid "as we will see" ... since it is rather "as the educated reader will see" ...
% }

As for photons, electron quantum optics started by probing the discrete nature of the electrons.
The granularity of electrons was undermined by the quantum fluctuations of the current \cite{Lesovik_1989} -- the so-called shot noise -- similar to photon noise \cite{Buettiker_1992,Landauer_1991}.
The first experiments mimicking textbook experiments of quantum optics, but with electrons, date back to 1999 with electronic Hanbury Brown and Twiss experiments~\cite{henny_99,oliver_99,liu_1998}.
These achievements were succeeded by the realisation of the first electronic Mach--Zehnder interferometer \cite{ji_2003}, a quantum device that nicely shows up the wave nature of the electron and which is a key tool for qubit manipulation.
All of these pioneering experiments that fostered the idea of electronic quantum control for computing applications have however been performed applying a DC current -- meaning a continuous stream of billions of electrons.
Only in 2007 the invention of the first single-electron sources \cite{feve_2007} opened up the possibility of performing electron-quantum-optics experiments at the single-particle level.
This achievement triggered tremendous progress on electron quantum optics over the last two decades bringing various single-electron sources \cite{feve_2007,blumenthal_2007,hermelin_2011,mcneil_2011,dubois_2013} that reach now emission efficiencies larger than 99\% \cite{dubois_2013,takada_2019,Freise2020Mar,giblin_2012,Stein2017Feb} -- a value far superior to latest single-photon sources \cite{senellart_2017b, munch_2013, Tomm_2021, moreau_2001, thomas_2021}.
Besides emission, also single-electron detection has significantly progressed.
Whenever a single electron is captured on a sufficiently long timescale -- of the order of a microsecond or more -- also detection efficiencies well above 99\% are achieved \cite{takada_2019,Freise2020Mar}.
The efficient control on the single-particle level points out the large potential to exploit single flying electrons for quantum applications.

From measurements on charge qubits in stationary double quantum dots \cite{hayashi_2003,petta_2004,petersson_2010,stockklauser_2017}, the coherence time of electron flying qubits is expected to be of the order of \he{a few nanoseconds}.
% several nanoseconds. \cbc{probably we should write of the order of one nanosecond; relaxation id of the order of 10 ns but coherence is much shorter}
Decoherence is, thus, the major obstacle for quantum implementations with single flying electrons.
The young research field of electron quantum optics is however constantly producing new findings bringing the coherence properties of flying electrons closer to macroscopic scales \cite{roulleau_2008, yamamoto_2012,duprez_prx_2019,duprez_science_2019}.
From the quantum-computation perspective, the central figure of merit is the number of operations that can be performed within the coherence time.
At present, this number is about 1000 for leading approaches such as superconducting qubits, trapped ions, or spin qubits~\cite{ladd_2010, ScienceNews2016, SC_qubit_review_2019, ion_review_2019, stano_2021}.
In order to achieve such operation fidelity with electron flying qubits, a time control at the picosecond scale is required.
Such ultrafast in-flight manipulation is presently being pursued in the FET-Open project {\it UltraFastNano} \cite{ufn_2020}.

The EU project {\it UltraFastNano}\footnote{www.ultrafastnano.eu -- https://cordis.europa.eu/project/id/862683}
aims to advance ultrafast operations in nanoelectronic devices to demonstrate the first electron flying qubit.
The concept is similar to that of a photonic quantum computer but, instead of photons, electrons are used as carrier of quantum information.
The project is a concrete example where the academic and the industrial sector are joining forces to develop and benchmark the tools that are required by the value chain of emerging quantum industry.
Often, the initial market of tools for sufficiently advanced technologies is too small to be pursued by larger corporations, opening a niche-market opportunity for small and medium-sized enterprises (SME).
This demand is satisfied by the industrial partner {\it nextnano GmbH} \footnote{www.nextnano.com} -- an SME based in Munich (Germany) with 12 employees and 300 customers in more than 35 countries -- which develops a software tool calculating the quantum mechanical properties of semiconductor nanodevices.
The synergy of academic and industrial partners on numeric simulations and experimental implementations within {\it UltraFastNano} fosters progress on electron flying qubits opening a novel branch of quantum industry.

In this article, we review three promising experimental routes towards electron-flying-qubit implementations and discuss the potential of numerical simulations to speed up experimental development cycles towards quantum-computing applications.
The reviewed experimental pathways differ mainly in the way the electron qubit is transported.
In particular, we address single-electron transport by means of a surface-acoustic wave, emission form a quantum-dot pump and Levitons.
For each transport approach, we point out the specific aspects where numerical simulations are key to unveil efficient routes for follow-up implementations.
Identifying these target aspects of cutting-edge electron-quantum-optics experiments, we finally present generic numerical simulations providing insights that are decisive for the development stages towards electron flying qubits.
Confronting numerical simulations with latest experimental results, we point out the capability of the numerical simulations to guide experimental implementations faster to success.

\section{Pathways towards the electron flying qubit}

Different than for a classical bit where the states 0 and 1 basically correspond to the charging state of a capacitor, the electron flying qubit is defined via the presence of a single electron in two paths of transportation.
The quantum state of the thus-defined flying qubit can be depicted on a Bloch sphere as shown in Fig.~\ref{fig:bloch}a.
The north and south pole of the sphere represent the classical states of the electron being in one of the two paths of transportation (0 and 1).
The probability to end up in one of these states is \he{represented}
% \sout{determined}
via angular coordinates of the sphere $\theta$ and $\phi$ that make up the quantum state $\psi$:
\begin{equation}
\label{RotState}
\ket{\psi} \propto\cos \left(\frac{\theta}{2}\right)\ket{0}+ \sin \left(\frac{\theta}{2}\right)\cdot\exp \left(i \phi \right)\ket{1}
\end{equation}

The thus-defined quantum state of the flying electron is fully controllable in an electronic Mach--Zehnder interferometry setup as sketched in Fig.~\ref{fig:bloch}b.
The quantum interferometer hosts two regions where the paths of transportation approach each-other and couple via a narrow potential barrier (see horizontal lines).
According to the potential landscape within this coupling-region -- see Fig.~\ref{fig:bloch}c --, the flying qubit state undergoes periodic oscillations (angle $\theta$) caused by coherent tunneling.
In between the two tunnel-coupled regions, the flying qubit picks-up a quantum phase $\phi$ that is tunable via the potential along the
% \sout{two-paths}
\he{two paths}
and the enclosed magnetic flux.
Employing a surface-gate defined quantum interferometer realised in a GaAs-based heterostructure (see Fig.~\ref{fig:bloch}d), basic quantum operations of such an electronic flying qubit have already been successfully demonstrated (see Fig.~\ref{fig:bloch}e) with a continuous current of electrons \cite{yamamoto_2012}.
The big challenge ahead is to perform such quantum state control on the level of single electrons and to couple several of the so-obtained electron flying qubits to generate a set of non-local entangled quantum states.

In semiconductor devices, a single electron is manipulable in a surface-gate-defined nanoscale structure such as a quantum dot or a waveguide.
The majority of such implementations are performed within the two-dimensional electron gas (2DEG) formed near the surface of a GaAs-based heterostructure \cite{bauerle_2018}, which have typically coherence lengths of several tens of micrometers \cite{niimi_2009, niimi_2010, yamamoto_2012, roulleau_2008}.
%
% [niimi_2009]
% Niimi Y, Baines Y, Capron T, Mailly D, Lo F Y, Wieck A D, Meunier T, Saminadayar L and Bäuerle C 2009 Effect of disorder on the quantum coherence in mesoscopic wires Phys. Rev. Lett. 102 226801
%
% [niimi_2009]
% Niimi Y, Baines Y, Capron T, Mailly D, Lo F Y, Wieck A D, Meunier T, Saminadayar L and Bäuerle C 2010 Quantum coherence at low temperatures in mesoscopic systems: effect of disorder Phys. Rev. B 81 245306
%
Applying a set of negative voltages on these gates, one can shape the potential landscape in the 2DEG and thus form and control the nanoscale devices.
\he{
	So far, there are two methods to guide the electrons along the desired paths:
	electrostatic waveguides \cite{yamamoto_2012, roussely_2018} or -- at high magnetic field -- quantum-Hall edge states \cite{ji_2003,bocquillon_2013}.
}
% \cb{To guide the electrons on the desired trajectories two possibilities are available.
% At low magnetic field electrostatic waveguides can be conveniently used \cite{yamamoto_2012, roussely_2018} while at high magnetic field the edge states in the quantum Hall regime can be exploited \cite{ji_2003,bocquillon_2013}.
% }
The 2DEG quality has been continuously improved \cite{thomas_2018,chung_2021} \he{and combined with clever device design \cite{roulleau_2008}, which allowed to} push the phase coherence length up to several hundreds of micrometers in recent experiments \cite{duprez_prx_2019}.

\he{
	The availability of quantum dots serving as highly efficient single-electron sources and receivers led to the development of single-electron-transport techniques based on surface acoustic waves \cite{talyanskii_1997,hermelin_2011,mcneil_2011} and voltage-modulation pumping \cite{kaestner_2008,yamahata_2017}.
	Fostered by technological progress and a growing understanding of the underlying physical mechanisms \cite{kaestner_2015}, these approaches now achieve transfer efficiencies well above 99\% \cite{takada_2019,Freise2020Mar,giblin_2012,Stein2017Feb}.
	The quantum-dot-based transport systems represent the electronic counterpart to the deterministic single-photon source we have mentioned earlier.
	Besides that, different avenues have been explored such as the so-called Leviton that is a well-protected collective single-electron excitation generated by an ultrafast Lorentzian voltage pulse \cite{dubois_2013, jullien_2014,bisognin_2019}.
	As the aforementioned photonic squeezed state that is a special kind of a coherent laser pulse, a Leviton represents a special form of a classical voltage pulse.
	The progress in these experimental routes -- surface acoustic waves, electron pumps and Levitons -- opened up the way to realise a flying-qubit platform with electrons instead of photons.
	In the following sections, we outline these three conceptually different approaches towards the electron flying qubit in more detail and discuss how numerical simulations have played a key role to interpret the experimental results guiding nanodevice design to the next generation.
}
% \cbc{very good! I like the description above}

% \todo{H.E.: There is room for improvement in the previous paragraph. The flow of arguments is yet not optimal ... but I would be also fine keeping it as it is.}

\subsection{Electron qubits surfing on a sound wave}
\label{sec:SAW}
\begin{figure}[!h]
	\centering
	\includegraphics[width=9cm]{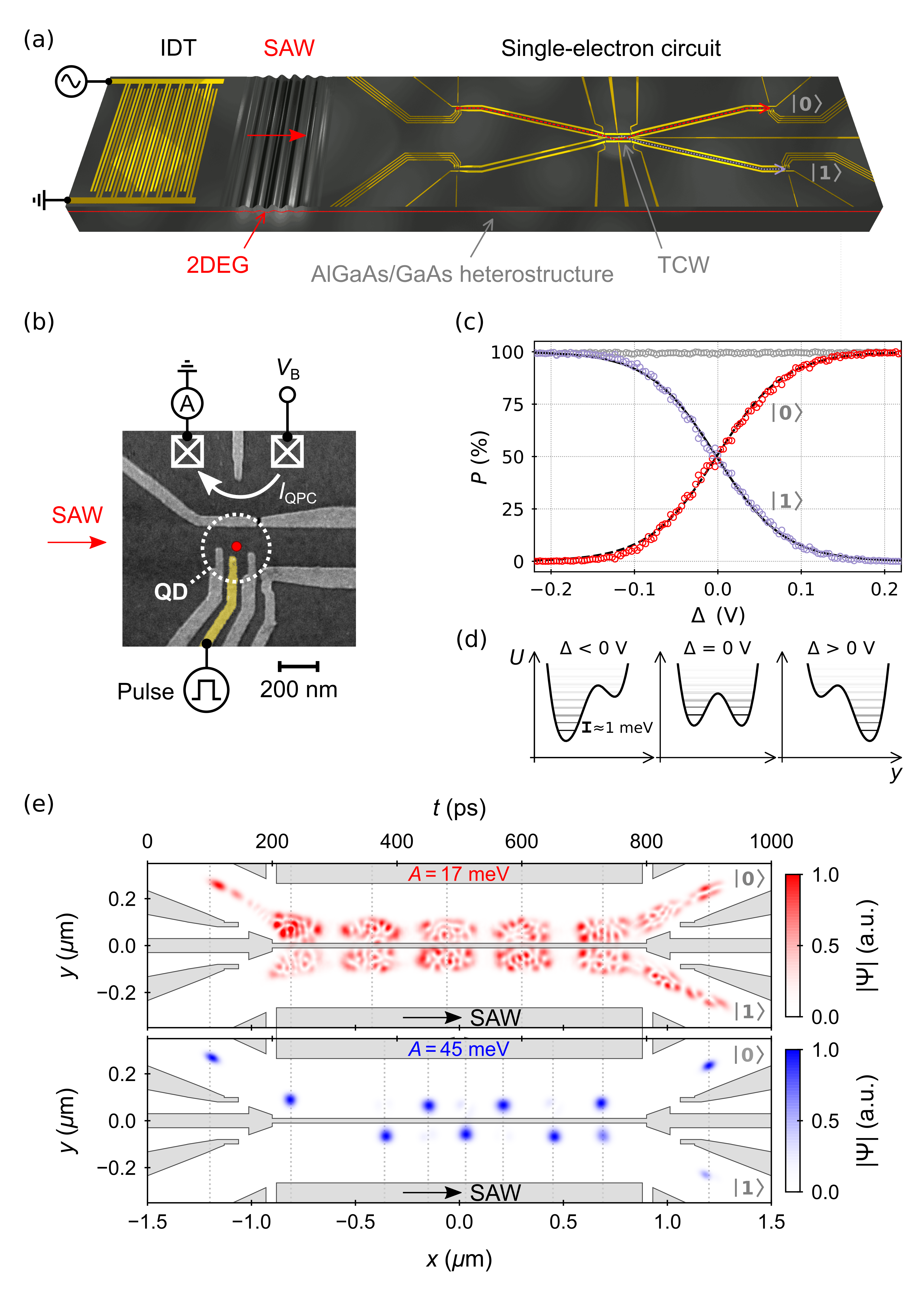}
	\caption{
		\textsf{\textbf{SAW-driven single-electron transport through a tunnel-coupled wire (TCW).}}
		\textsf{\textbf{(a)}} Schematic showing \he{an} interdigital transducer launching a surface acoustic wave (SAW) towards the circuit consisting of a pair of tunnel-coupled transport paths whose ends are equipped with QDs to send and catch a single electron.
		\textsf{\textbf{(b)}}
		Scanning electron micrograph of the source quantum dot (QD) with schematic indications of the QPC sensor and the pulsing gate.
		\textsf{\textbf{(c)}}
		Measurement of the single-shot transfer probability $P$ as function of potential detuning $\Delta$.
		The solid line shows the result of a simplistic one-dimensional simulation assuming charge excitation in the TCW.
		\textsf{\textbf{(d)}}
		Schematic showing vertical potential cuts in the tunnel-coupled wire (TCW) for different potential detuning $\Delta$.
		\textsf{\textbf{(e)}}
		Time-dependent simulation of SAW-driven electron propagation through the tunnel-coupled wire (TCW) at $\Delta=0$ for peak-to-peak SAW modulation amplitudes of $A=\SI{17}{meV}$ and $A=\SI{45}{meV}$.
		The grey regions indicate the surface gates.
		The data shows the evolution of the single-electron wave function $\psi$ for different time frames $t$ indicated via the vertical dashed lines. Adapted from Ref.~\cite{takada_2019}.
	}\label{fig:saw}
\end{figure}

In III-V
%\OYe{explain?}\he{HE: The crystal structures have no inversion symmetry, therefore piezoelectric. I think it would go a bit to much in detail.}
semiconductor compounds, such as the presently discussed GaAs-based devices, sound is accompanied by an electric potential wave due to its piezoelectric properties allowing charge displacement \cite{Wixforth1986}.
At the first glance, acousto-electric transport seems a rather brute approach to move an electron qubit.
A surface acoustic wave (SAW) has proven itself however as \he{an} efficient and well-controllable transport medium.
Fig.~\ref{fig:saw}a shows a schematic of the single-electron transport approach.
A SAW is typically generated with an interdigital transducer (IDT) -- a device that is well established in modern consumer electronics products~\cite{morgan_2007}.
Applying a finite, resonant input signal on the IDT, a SAW is emitted which then travels relatively slowly with a characteristic speed of about \SI{3}{\micro\meter\per\nano\second} towards a surface-gate-defined nanoscale device \cite{delima_2005}.
The circuit is constructed on the basis of fully depleted transport channels whose ends are equipped with quantum dots (QD) -- see Fig.~\ref{fig:saw}b -- serving as highly efficient single-electron source and receiver.
Each QD is equipped with an adjacent quantum point contact (QPC) allowing to trace its charge occupation.
After loading a single electron at the source QD via a sequence of voltage variations on the corresponding surface gates, a SAW is emitted.
The SAW train typically has a duration of tens of nanoseconds and wavelength of \SI{1}{\micro\meter}.
When arriving at the depleted transport channel, the potential modulation of the SAW forms a train of moving quantum dots propagating through the surface-gate-defined rail.
After loading an electron at the source QD, this SAW train allows to shuttle the electron through the quantum rail to a distant receiver QD \cite{hermelin_2011,mcneil_2011}.

The robustness of a SAW train enables acousto-electric transfer of a single electron in a nanoscale circuit approaching macroscopic dimensions.
An experimental investigation of a \SIadj{22}{\micro\meter}-long SAW-driven single-electron circuit consisting of two tunnel-coupled channels -- see single-electron circuit in Fig.~\ref{fig:saw}a -- achieved single-shot transfer efficiencies larger than 99\% \cite{takada_2019}.
Here, the exact sending position within the SAW train is controlled via the delay of a picosecond-scale voltage-pulse trigger applied on the source QD.
Adjusting the potential landscape in the tunnel-coupled wire (TCW), it is possible to partition the electron wave function via directional coupling at will between the two transport channels.
Fig.~\ref{fig:saw}c shows an exemplary measurement of the single-shot transfer probability $P$ as function of potential detuning $\Delta$ in the TCW.
The partitioning data shows a constantly high transfer efficiency despite the detuning $\Delta$ of the TCW potential.
The shape of the partitioning data bears important information on the time-evolution of the flying quantum state that is of central importance for quantum applications.

To draw the right conclusions from the single-shot partitioning data, numerical simulations are essential.
In the presently discussed example, time-dependent simulations of the electron's propagation through the TCW region, revealed charge excitation -- as schematically shown via the potential landscapes shown in Fig.~\ref{fig:saw}d -- due to insufficient SAW confinement.
In these calculations, the stationary potential is \he{calculated with the {\it nextnano} software} based on the true sample geometry and the electric properties of the heterostructure.
Superposing this potential landscape with the dynamic modulation of the SAW, one can prepare an electron in its ground state and simulate its propagation through the device as shown in Fig.~\ref{fig:saw}e.
Setting a SAW-modulation as present in the experiment ($A \approx \SI{17}{\milli eV}$), the simulation shows a picture that is in good agreement with the experimental observation:
As the electron enters the TCW, insufficient confinement within the moving, acousto-electric QD provokes charge excitation that prevents the appearance of tunnel oscillations.
Instead, the probability to find the electron spreads according to the excitation spectrum.
Unlike the experiment, numerical simulations allow to examine the effect of various device parameters in a systematic and fast way.
For the presently discussed example, the time-dependent simulations particularly showed up the importance of the acousto-electric in-flight confinement.
Augmenting the SAW amplitude by a factor of three ($A \approx \SI{45}{\milli eV}$), the time-dependent simulations predict preservation of quantum confinement at TCW transit resulting in tunnel oscillations \cite{takada_2019}.
Since an increase of the acousto-electric power of this scale is technically feasible, the time-dependent quantum simulation points out a central, easily addressable aspect in the realisation of electron flying qubits transported by sound.

Following the numerically guided pathway of augmented acousto-electric amplitude, we anticipate coherent in-flight manipulations of flying charge qubits on technically relevant length scales soon in single-shot experiments.
For a flying qubit employing the electron's charge, first observations of tunnel-related probability oscillations have been already reported from experimental studies on SAW-driven transport of a continuous stream of single electrons  \cite{kataoka_2009,ito_2021}.
In congruency with the prediction of the aforementioned time-dependent simulations, the threshold of the SAW amplitude to significantly confine an electron in a single acousto-electric minimum was recently determined as $A=(24\pm3)\;\textrm{meV}$ in flight-time measurements \cite{edlbauer_2021}.
For electron flying qubits defined by spin, increased SAW amplitude have already helped to demonstrate coherent transport of an entangled electron pair over $\SI{6}{\micro\meter}$ distance \cite{jadot_2021}.
The coherent acousto-electric transfer of  a single electron between remote quantum dots marks a new route to link quantum information in semiconductor qubit circuits where numerical simulations will certainly play a central role to identify key aspects and speed up experimental cycles.

\subsection{Hot-ballistic electrons}

On the contrary to the aforementioned acousto-electric transport approach, a lithographically-defined, but highly-tunable, quantum dot can also be employed to emit a single electron at high energy.
These hot-electron sources for the controlled emission of single and multiple particles are of high interest from the perspective of higher-temperature operation and isolation from environment.
In these devices, electrons can be emitted at an energy $\sim 100$~meV above the Fermi energy, hence the cooling of the Fermi sea at millikelvin temperatures may not be necessary.
Besides, hot electrons can be transmitted through a depleted channel, eliminating undesirable interactions with the Fermi sea.
\he{For the controlled emission of single and multiple particles, hot-electron sources are driven by strong potential modulation determining the timing of electron emission via slow, stochastic tunneling  through a barrier.}
% \sout{For the controlled emission of single and multiple particles, hot-electron sources are driven by strong potential modulations, by which the timing of electron emission is determined, rather than by a stochastic process through a slow tunnelling barrier.}
This \he{process} has a potential advantage of a high purity, meaning that the energy and time window into which the particles (or wave packets) are emitted fluctuates little between
successive emissions.
% \todo{M.K.: We explain in a few sentences, how the electron degrees of freedom cause inelastic scattering and decoherence -- also in respect to the other approaches mentioned in the manuscript.}
On the other hand, due to a large phase space available, the inelastic scattering rate during propagation can be high, leading to short decoherence time.

\he{
	This is in contrast to the electrons confined in SAW potential or the Levitons, for which the electrostatic confinement or the limitation in available states (by the filled Fermi sea), respectively, protect the states from scattering processes.
	The nature of inelastic processes for hot electrons injected in GaAs/AlGaAs heterostructures has been investigated in Ref.~\cite{Taubert2010,Taubert2011,Fletcher2013,Emary2016Jan,Emary2019Jan,Johnson2018Sep,Ota2019}.
	At zero or small magnetic fields, the dominant scattering process is electron--electron interactions by the Fermi sea.
	For low-energy electrons (a few tens of meV above the Fermi energy), the electron--electron interactions continue to be the dominant process at higher magnetic fields applied \he{in} perpendicular to the plane of two-dimensional electron gas.
	For high-energy electrons ($\sim 100$~meV above the Fermi energy), the magnetic confinement to the channel edge limits the spatial overlap with the Fermi sea and consequently suppresses electron--electron interactions.
	Instead, the emission of longitudinal optical phonons \cite{Heiblum1985} becomes the dominant process.
	Generally, the optical-phonon emission rate at high magnetic fields tends to be smaller than the electron--electron scattering rate at low magnetic fields, and therefore the ballistic transport length tends to be larger at higher magnetic fields.
	The suppression of backscattering
	\he{processes}
	% process
	due to chiral transport also contributes to a longer transport length at higher fields.
}

The technology to use hot electrons for electron quantum optics
\he{experiments}
% experiment
is relatively new and not much information has been gathered regarding their suitability for applications in quantum information processing.
In order to explore the potential to use these hot-electron sources for the preparation of flying qubits, it is important to gain very precise knowledge of the relevant properties of the injected particles.
These are (i) the time- and energy interval into which particles are emitted (ii) the purity of the wave packets, namely the precision in the time- and energy-interval in which the particle is detected in every driving cycle.
At the same time it is crucial to analyse, (iii) how these properties are affected during the propagation of wave packets along depleted channels and to minimise a possible deterioration of the signal.
These aspects hence need to be tested and control over them has to be obtained.

\begin{figure}[!h]
	\centering
	\includegraphics[width=9cm]{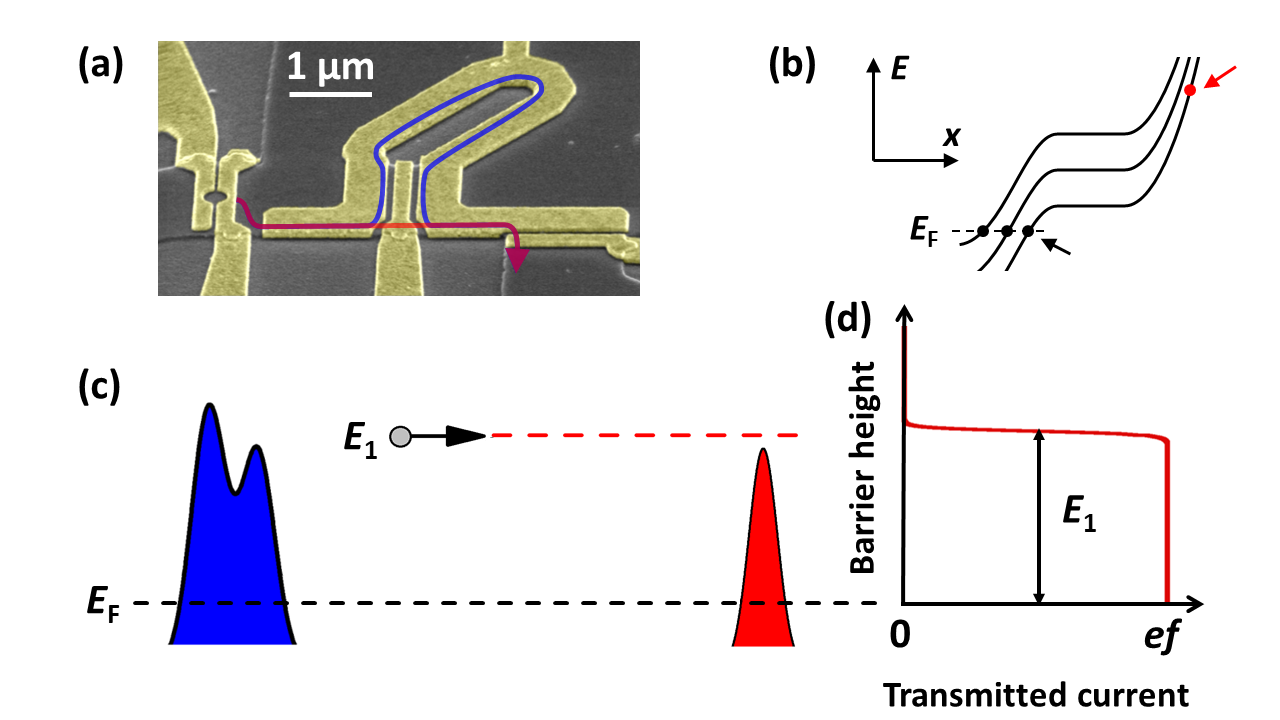}
	\caption{
		\textsf{\textbf{Hot-electron emission and transport.}}
		\textsf{\textbf{(a)}} A false-colour SEM image of a hot-electron time-of-flight measurement device \cite{Kataoka2016Mar}.
		The red and blue arrows
		\he{show two paths taken by the electrons emitted by a quantum-dot pump at the left.}
		% show two electron paths emitted by an electron pump shown on the left.
		The difference in arrival times from the two paths are used to estimate the drift velocity.
		The two-dimensional electron gas in the paths are depleted by a surface gate.
		\textsf{\textbf{(b)}}
		A schematic of Landau levels near the edge where \he{the} two-dimensional electron gas is depleted.
		``Usual'' edge states forming near the Fermi energy are indicated by black circles.
		The high-energy states where hot electrons travel are indicated by a red circle.
		\textsf{\textbf{(c)}}
		Energy diagram of hot-electron emission from an electron pump (blue-coloured potential on the left) and its detection by energy-dependent barrier (red-coloured potential on the right).
		(d) The determination of the electron emission energy $E_{\rm{1}}$ by the measurement of transmitted current as a function of the detector barrier height \cite{Fletcher2013}.
		$f$ is the operating frequency of the electron pump.
	}\label{fig:Hot_electron_emission}
\end{figure}

The detection of hot-ballistic electrons emitted by a single-electron source was made using a scheme shown in Fig.~\ref{fig:Hot_electron_emission}~\cite{Fletcher2013}.
The energy distribution of hot electrons was obtained from the transmitted current through a detector barrier.
In addition to the main distribution around the emission energy, replica of the distribution with discrete energy steps $\sim 36$~meV were experimentally observed.
They were attributed to the emission of longitudinal-optical phonons.
Further studies on phonon interactions~\cite{Emary2016Jan,Emary2019Jan} led to a method to suppress phonon emission probabilities by softening the edge potential~\cite{Johnson2018Sep,fujisawa_2019}.
This technique was used to extend the phonon scattering length to as much as $\sim 1$~mm.  Using the long ballistic length, time-of-flight measurements were performed to extract the electron drift velocity ranging from 30 to \SI{130}{\micro\meter\per\nano\second}~\cite{Kataoka2016Mar}.

\begin{figure}[!h]
	\centering
	\includegraphics[width=7cm]{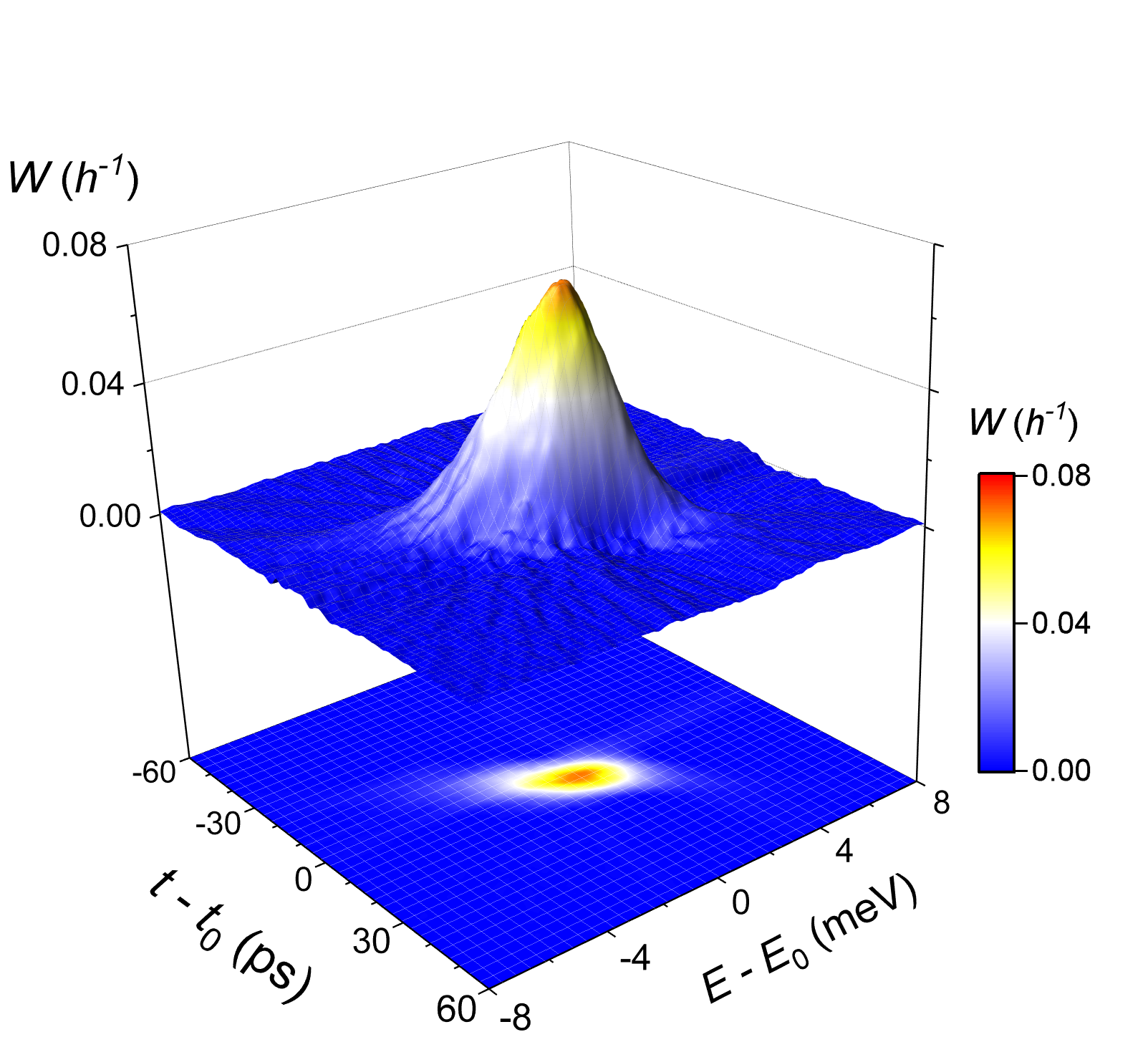}
	\caption{
		\textsf{\textbf{Hot-electron quasiprobability distribution.}}
		% A
		Wigner quasiprobability distribution (a map to visualise the particle's state in phase space by translating the wave function) plotted on the energy ($E$)-time ($t$) phase space using \he{the} time-dependent barrier described in Ref.~\cite{Fletcher2019Nov}.
		$E_{\rm{0}}$ and $t_{\rm{0}}$ are
		% arbitrary
		\he{arbitrarily}
		chosen central values of \he{the} electron emission energy and arrival time.
		This plot shows that the quasi-probability distribution has a correlation between energy and time, implying that the emission energy is lifted as the electron leaves the source.
	}\label{fig:Hot_electron_Wigner}
\end{figure}

The time-of-flight measurements used a time-gating technique to measure the electron arrival-time distribution \cite{Waldie2015Sep}.
This was later developed into a tomographic measurement of quasi-probability distribution in the energy-time phase space by controlling the ramp speed of gate voltage \cite{Kataoka2017Mar,Fletcher2019Nov,Locane2019Sep}.
This measurement revealed an energy-time correlation of the distribution imprinted by the ramp speed of source energy state during the emission process (see Fig.~\ref{fig:Hot_electron_Wigner}) \cite{Fletcher2019Nov}.
The projection of this distribution onto the time or energy axis gives the arrival-time or energy distribution of the emitted states.
The purity of the observed distribution was only 0.04, and therefore the observed state is likely to be a mixed state. We note that the time and energy resolutions of the experiment in Ref.~\cite{Fletcher2019Nov} were estimated to be $\sigma_{t} \simeq 0.3$~ps and $\sigma_{E} \simeq 0.8$~meV, giving $\sigma_{t} \sigma_{E} \simeq 0.36 \hbar$, implying that this method is capable of resolving the minimum uncertainty limit ($\hbar / 2$).  Therefore the observed low purity is not due to poor measurement resolutions, but is likely due to noises in electron emission process from the source.
In this set of experiment, the smallest arrival-time distribution observed was $\sigma_{t,\rm{min}} \simeq 5$~ps, and the smallest product of time and energy widths was $\sim 30$ times larger than the minimum uncertainty limit (taking into account the energy-time correlation).
A method has been proposed in Ref.~ \cite{Ryu2016Sep} to emit each electron into Gaussian-shaped minimum uncertainty states.
Another important experimental technique is the full counting statistics of the electron number partitioned by a beam splitter (a tunnel barrier).
This has been demonstrated using noise measurements \cite{Ubbelohde2015Jan} or a trap coupled to a single-charge detector \he{\cite{fricke_2013,Freise2020Mar}}.

The time scale of electron emission directly reflects in the width of the emitted wave packets. In quantum optics experiments with electrons, this time scale is important for the visibility of interference effects.
In order to obtain these insights, the times at which single or multiple particles are emitted from the hot-electron sources can be studied analytically or numerically taking into account the time-dependent modulation of the single-particle energy levels, as well as of the shape of the tunneling barrier between quantum dot and conductor~\cite{Waldie2015Sep}.
Importantly, theoretical studies have recently shown that Coulomb interaction between electrons on the dynamically driven quantum dot have a strong impact on the energies at which the particles are emitted. Most crucially this impact on the electron energies also directly influences the time scale on which the emission of the different particles takes place~\cite{Schulenborg2016Feb}.
This theory furthermore predicts that the separation of time scales becomes particularly relevant for energy-dependent barriers~\cite{Schulenborg2018Dec}.
Different schemes of how to read out these different relevant \textit{emission} time scales using side-coupled detector dots~\cite{Schulenborg2014May} or nonadiabatic pumping schemes~\cite{Riwar2016Jun} have been suggested.
This last scheme in particular also addresses relaxation times due to phonons during the \textit{emission} process.

Based on the work described above, one can expect that realisations of Mach--Zehnder experiments will become possible with these types of sources, as suggested in Ref.~\cite{Clark2020Oct,Barratt2021Apr}, similar to previous proposals for single- and two-particle interferometers for minimal-excitation single-particle sources~\cite{Haack2011Aug,Juergens2011Oct,Splettstoesser2009Aug}.
Ref.~\cite{Barratt2021Apr} studied phase-averaging effects, which are particularly important for the temporally-short, high-energy single-electron
wave packets.
As a result it becomes necessary to tune asymmetric interference arm lengths and delay time, which could be achieved by tuning the drift velocity.
These analytical studies~\cite{Clark2020Oct,Barratt2021Apr,Haack2011Aug,Juergens2011Oct,Splettstoesser2009Aug} assume an emission of pure states and ideal beam splitters, which are over-simplified compared to a realistic experiment. Hence, in order to improve the device characteristics, more realistic numerical modelling of these aspects could be a helpful complement.
While the electron coherence of hot electrons is yet to be demonstrated, the short length of electron wave packet in time domain and the ability to control their emission timing with a picosecond resolution can be useful in ultrafast electronics applications.
In-situ voltage sampling under cryogenic environment has been demonstrated with a bandwidth potentially exceeding 100~GHz \cite{Johnson2017Mar}.
This technique was used to determine the precise gate voltage ramp profile for quantum tomography measurements \cite{Fletcher2019Nov}.

\subsection{Leviton qubits flying over the Fermi sea}

\begin{figure}[!h]
	\centering
	\includegraphics[width=9cm]{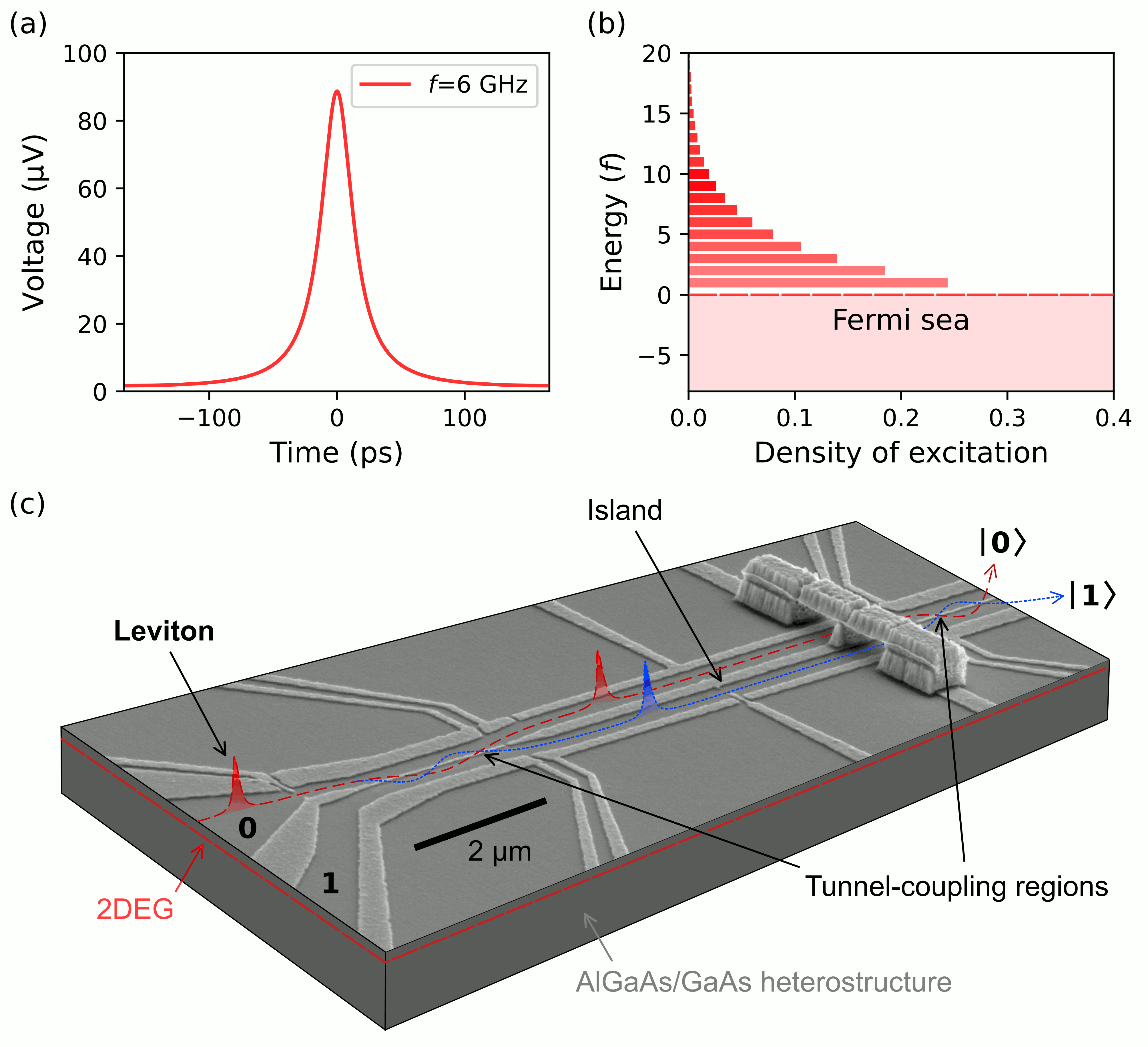}
	\caption{
		\textsf{\textbf{Levitons qubits.}}
		\textsf{\textbf{(a)}}
		\he{Time trace of a Lorentzian voltage pulse that is suitable for excitation of Levitons.}
		%   Time trace of a Lorentzian voltage pulse of temporal width of 20 ps that is suitable for Leviton excitation.
		\textsf{\textbf{(b)}}
		Excitation spectrum corresponding to this Lorentzian pulse showing the collective excitation of the single-electron wave function without perturbation of the Fermi sea.
		\textsf{\textbf{(c)}}
		Scanning electron micrograph showing a quantum interferometer with the two
		\he{tunnel-coupled wires enclosing the}
		%   tunnel-coupling regions and the enclosed
		island with \he{schematic} indication of the paths of \he{for transportation of the Levitons}.
		%   \cbc{we should also put kets for the injection rails}
		%   \comment{H.E.: I think it is nice like this because it implicitly shows that we start with a classical state (without kets) that we make quantum (with kets) by passing it through the interferometer -- compare Fig.1.}
	}\label{fig:leviton}
\end{figure}

A conceptually different approach to realise \he{an} electron flying qubit is to generate a single-electron wave packet directly from the Fermi sea.
This approach seems
\he{counterintuitive}
% to be quite counter intuitive
as \he{a} perturbation of the Fermi sea excites both electrons and holes and does in principle not allow the generation of a \textit{pure} single-electron wave packet.
L. S. Levitov and co-workers came up with an original idea
\he{to form a collective electron excitation flying over the Fermi sea without leaving a hole \cite{Levitov_1996,Ivanov_1997,Keeling_2006}.}
% to collectively form a single electron flying without excitation of holes over the Fermi sea \cite{Levitov_1996,Ivanov_1997,Keeling_2006}.
% \OYe{"collectively" does not sound
% good here, might be related to many co-authors :-) I'd reformulate}
It has been shown that a voltage pulse of Lorentzian shape:
\begin{equation}
V(t)= \frac{V_0}{\pi}\frac{\tau/2}{t^2+(\tau/2)^2}
\label{eqn:lorentzian}
\end{equation}
% \sout{can generate}
\he{generates}
a \textit{pure} single-electron excitation -- the so-called Leviton -- if the amplitude $V_0$ and the full-width at half-maximum $\tau$ are chosen to match the quantization condition:
\begin{equation}
\int_{-\infty}^{\infty}eV(t)\;\textrm{d}t = h
\label{eqn: quantization}
\end{equation}
where $e$ is the elementary charge and $h$ is Planck's constant.
A Lorentzian pulse fulfilling this quantization condition is shown in Fig.~\ref{fig:leviton}a.
Fig.~\ref{fig:leviton}b shows the corresponding excitation spectrum -- meaning the occupation of states above and below the Fermi energy.
The calculation shows a distribution that is characteristic for Leviton excitation.
The collective wave function is only occupying the states right above the Fermi level (zero energy) forming a \textit{pure} electronic excitation that is robust against relaxation.

It took almost 20 years until the theoretical concept of a Leviton was demonstrated in experiment \cite{dubois_2013}.
The reason for this long delay was mainly related to the
\he{difficulty in generating clean and sufficiently short voltage pulses}
% \sout{generation of the required ultra-short voltage pulses}
\he{of Lorentzian shape}
that are injected directly via an \he{O}hmic contact of the quantum device.
Compared to the aforementioned quantum-dot-based sources, the Leviton approach brings the advantage that nanolithography techniques are not required to define single-electron emitters.
At last, progress in microwave engineering has bridged this gap and allowed one to verify this original concept.
The experiment demonstrated minimization of shot noise due to the absence of holes via Leviton formation and Hong--Ou--Mandel type experiments with very high degree of indistinguishability.
To study the wave function of such flying charge excitation~\cite{Keeling_2006,Grenier_2013,Moskalets_2015,Glattli_2016,Ferraro_2018,Vanevic_2016,Yin_2019,flindt_2021}, quantum tomography protocols have been developed allowing a measurement of the Wigner distribution function~\cite{jullien_2014,bisognin_2019,Grenier_2011,Roussel_2021}.
In addition, time-resolved experiments have shown propagation of Leviton-like flying charges over distances of more than \SI{80}{\micro\meter} without measurable dispersion \cite{roussely_2018}.
Owing to the occupation of states right above the Fermi sea, Levitons are expected to have extremely good coherence properties \cite{ferraro_2013,ferraro_2014,Cabart_2018,Rebora_2020} compared to other injection schemes~\cite{Wahl_2014,freulon_2015,Marguerite_2016} making them highly promising candidates for electron-flying-qubit implementations.

The next important step to benchmark these benefits is the implementation of a quantum interferometer with Levitons.
Since Levitons are simply injected via voltage pulses on an ohmic contact, the geometry of such a single-qubit device is very similar to that of early experiments \cite{yamamoto_2012}. Figure~\ref{fig:leviton}c shows a SEM image of a possible implementation with schematic indications of the interferometer paths.
% \todo{H.E.: Describe that short pulses are important to have experimental control to study the dynamics in the coupling region -- eventually short pulses are not necessary.}
The propagation velocity of the injected Leviton pulse is expected to be on the order of \he{$100$ \SI{}{\micro\meter\per\nano\second}} \cite{roussely_2018}.
\he{
	Since the dynamics of such propagating pulses within a quantum interferometer have not been investigated yet, it is important to have experimental control of the pulse width to fit the flying charge excitation within the tunnel-coupling regions ($\approx\SI{2}{\micro\meter}$) requiring pulses with full-width-at-half-maximum smaller than \SI{20}{\pico\second}.
}
% \sout{To fit the flying charge excitation within the tunnel-coupling regions ($\approx$\SI{2}{\micro\meter}) of the quantum interferometer, it is thus necessary to generate Lorentzian voltage pulses having a full-width-at-half-maximum smaller than \SI{20}{\pico\second}.}
The generation of such pulses with cutting-edge microwave synthesis approaches is possible but at the technical limit.
A promising alternative route allowing pulses with widths of \SI{1}{\pico\second} or smaller is the optoelectronic generation via ultrafast photo switches \cite{Auston1975, Mourou1981, Auston1984, heshmat2012, giorgos_2020}.
Besides a proper Leviton source, it is of utmost importance to design a quantum interferometer structure allowing for qubit manipulation with maximum efficiency.
For this purpose, numerical simulations
% \sout{can serve}
\he{serve}
as a useful tool to model the evolution of quantum states along the interferometer structure.
In order to deduce the coherence length of a certain implementation it is necessary to measure the strength of the quantum oscillations for devices with successively increasing island-length.
The knowledge on these aspects of a single electron flying qubit made up by Levitons will be decisive for the applicability in quantum-computing implementations.

% \todo{Preden and Christian: Small paragraph comparing single-electron transport in quantum Hall edge channels to waveguides at low B field.}
\he{
	% Flying qubits based on Levitons have been implemented in electronic waveguides made with split-wires realized by patterning a 2D electron conductors confined in a GaAs/GaAlAs heterojunctions.
	% Another platform, also based on 2D electron systems, can be realized when putting the conductor in very high magnetic field to reach the quantum Hall
	% % \sout{Effect}
	% regime.
	% The 2D conductor can be obtained in lightly doped Graphene or in GaAs/GaAlAs heterostructure.
	% In this regime the conductor becomes topologically insulating and the only quantum channels carrying the current occur on the sample edge.
	% The propagation being chiral the quantum edge channel offers very long mean free path and high coherence.
	% Quantum Point Contacts (QPCs) can be used to realize electronic beam-splitter.
	% By combining two QPCs, electronic Mach--Zehnder interferometer can be realized \cite{roulleau_2007} to perform all qubit operation on the Bloch sphere.
	One route to realise flying qubits based on Levitons employs electronic waveguides defined by surface gates in the two-dimensional electron gas (2DEG)
	% }
	% \oy{2DEG} \OYe{\sout{the two-dimensional electron gas (2DEG)} -- 2DEG was introduced already}
	% \he{H.E.: It is true that it was already introduced. However, the abbreviation was used pages before the last time. In order to not exclude the uneducated readers, it is beneficial to mention it one more time. It does not harm.}
	% \he{
	of a GaAs/GaAlAs heterostructure \cite{yamamoto_2012,roussely_2018}.
	An alternative platform is transport along quantum Hall edge channels \cite{ji_2003, roulleau_2008}.
	It is formed when a 2DEG is placed in a very large magnetic field.
	In this regime the bulk becomes insulating
	% }
	% \OYe{better: "insulating and topologically nontrivial", "topologically insulating" sounds like jargon}
	% \he{H.E.: Indeed. I removed the "topologically", but skipped the "topologically nontrivial", since it is maybe not so relevant in the context.}
	% \he{
	and the only quantum channels carrying the current occur on the sample edge.
	It is applicable in the aforementioned GaAs framework or lightly doped graphene.
	The implementation of quantum point contacts (QPCs) enables the realisation of an electronic beam-splitter.
	By combining two QPCs, an electronic Mach--Zehnder interferometer is realised \cite{ji_2003,roulleau_2007} allowing full qubit manipulation on the Bloch sphere.
	Being chiral, the propagation along quantum edge channels offers a very long mean free path and coherent transport \cite{duprez_prx_2019}.
	In this regime, all the electron quantum optics tools are realizable such as Hanbury-Brown--Twiss \cite{neder_2007} and Hong--Ou--Mandel interferometry \cite{bocquillon_2013}.
	Single electron sources based on Levitons have been also implemented, particularly in graphene \cite{forrester_2014,forrester_2015}.
	Compared to quantum wires based on electronic waveguides at low magnetic field, the advantage is a
	% near
	\he{nearly}
	perfect free propagation of electrons, thanks to the chiral nature of
	% propagation
	\he{the edge channels}
	(electrons cannot go back after scattering on impurities).
	The drawback is the use of few Tesla magnetic fields and the chiral nature of the propagation which makes coupling of more than two electron flying qubits challenging.
}
% \OYe{Question: did anybody try to use topological insulators with SOI but no magnetic field?}

Single-shot detection represents a major
\he{task}
% \sout{aspect}
to realise an electron flying qubit with Levitons.
For the aforementioned investigation of quantum oscillations, statistical measurements are sufficient.
In order to control single Leviton qubits individually, it will be however necessary to
\he{detect}
% \sout{spot}
the presence for each flying electron via capacitive coupling to an ultra-sensitive quantum detector.
One possible implementation of such a quantum detector is a spin qubit that is operated in a regime where it is extremely sensitive to charge fluctuations \cite{thalineau_2014}.
At present, this type of detector is capable of sensing a few electrons and enables a quantum non-demolition measurement \cite{Meunier_2021}.
The quantum detector is able to record the presence of a passing flying electron without perturbing its quantum state that can in turn be reused after detection for further quantum manipulations.
This aspect is an important advantage over single-photon detectors where the photon disappears after detection.
Another possible implementation that has been put forward recently guides the flying electron through a meander structure which is capacitively coupled to two large metal electrodes.
The passage of the flying electron beneath the two surface electrodes generates an oscillating voltage signal.
This detector is expected to have a sub-electron sensitivity \cite{glattli_2020} and, when properly integrated into a quantum circuit, can also be adapted for quantum non-demolition measurements.

\begin{figure}[!h]
	\centering
	\includegraphics[width=9cm]{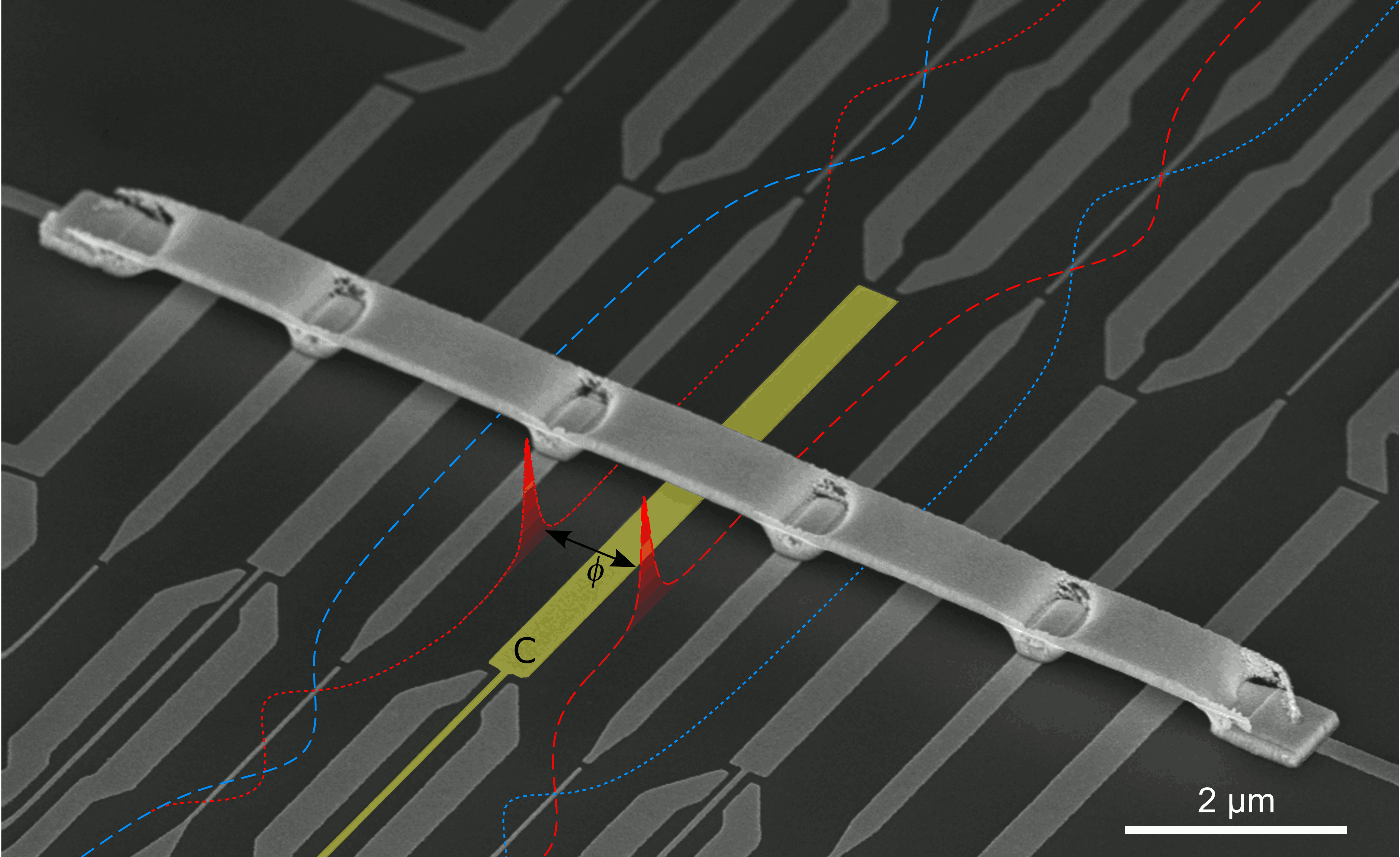}
	\caption{
		\textsf{\textbf{Scanning electron microscopy image of a multi-qubit flying electron architecture.}}\\
		The image shows four quantum interferometers that can be simultaneously operated owing to a common bridge that connects the islands of each device.
		The dashed lines schematically indicate the paths of two Levitons in two neighboring interferometers.
		The intermediate gate C (highlighted in yellow) allows for controlled Coulomb coupling of the Levitons and thus in-flight entanglement.
	}
	\label{fig:multi-qubit-SEM}
\end{figure}

Another aspect of major importance is the scalablilty of surface-gate defined quantum-interferometer devices.
Fig. \ref{fig:multi-qubit-SEM} shows a SEM image of a prototype multi-qubit implementation hosting four quantum interferometers.
Simultaneous operation of the electron flying qubits is accomplished via an extended bridge cross-connecting island-gate of each device.
To implement a two-qubit gate in such a setup the Coulomb interaction of flying electrons is exploited.
Let us consider the case where two Levitons are simultaneously sent through a pair of neighboring quantum interferometers.
By adjusting the potential barrier of a Coulomb-coupling gate (C) -- as shown in Fig.  \ref{fig:multi-qubit-SEM} -- the flying electrons are exposed to their respective Coulomb potential which introduces a quantum phase $\phi$ causing entanglement.
The phase induced by each of the two electrons is proportional to the coupling constant and the interaction time, hence to the gate length.
The Coulomb-coupling strength can be adjusted by changing the gate voltage on the electrostatic gates defining the phase-exchange window.
The coupling region can be as short a 1 µm for the case of ballistic electrons \cite{Ionicioiu_2001a, Bertoni_2000} and much shorter for the case of SAW-transported electrons since the propagation speed is 100 times smaller.
If a $\pi$ phase shift is induced, the probability of detecting an electron in the output port $\ket 0$ or $\ket 1$
is inverted, hence realising a controlled phase gate C$_\phi$.
Combining this experimental setup with 2 interferometers one can even go one step further and test Bell’s inequalities as proposed in reference \cite{bauerle_2018, Ionicioiu_2001b}.
In this case, all of the four beam splitters are used and a $\pi$ phase shift is induced with the Coulomb coupler enabling the formation of a maximally entangled Bell state.
The scalablilty of electron-flying-qubit implementations is very similar to that of photonic circuits where multiple Mach--Zehnder interferometers are connected in parallel and series \cite{slussarenkoa_review_2019}.
% \cbc{in this ref the DOI appears 2 times !} # H.E. solved

The central challenge to build a quantum computer is to scale up a qubit system.
For the latest technological stage, millions of \he{physical} qubits would be required \cite{Reiher_2017}.
This scalability problem is inherent to any qubit that needs to be addressed individually via an external parameter such as a gate voltage or a laser.
Important issues to be solved on the way to build a universal quantum computer are presently the improvements of the fidelity of the qubits as well as their connectivity \cite{Waintal_2020}.

\begin{figure}[!h]
	\centering
	\includegraphics[width=9cm]{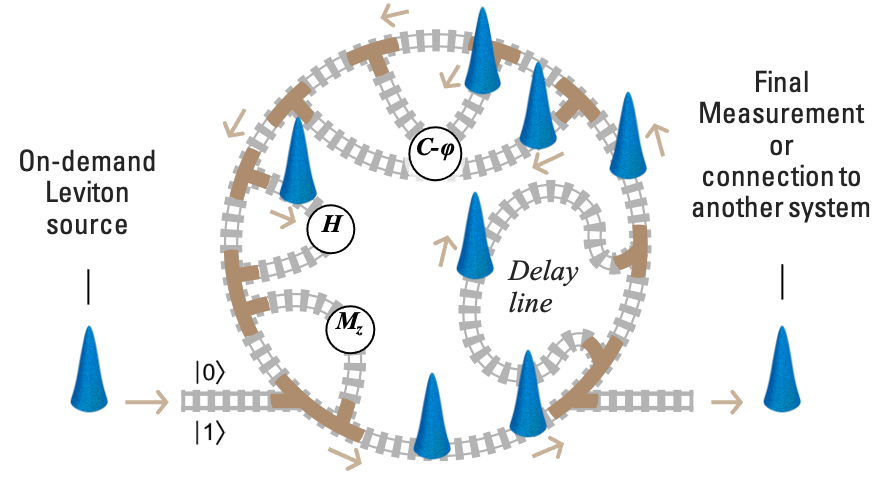}
	\caption{
		\textsf{\textbf{Architecture for a universal quantum computer using Levitons.}}
		The quantum switches (in brown) send the qubits to the various quantum gates. Single qubit rotations Hadamard (H), two-qubit controlled-phase (C-$\phi$) and measurement along z (M$_Z$) are implemented during the flight.
	}
	\label{fig:leviton-UCQ}
\end{figure}

Electron flying qubits using Levitons could allow one to implement an original architecture to build a universal quantum computer as schematized in Fig.~\ref{fig:leviton-UCQ}.
Although the architecture of Fig.~\ref{fig:leviton-UCQ} is a theorist view at this stage, it has very appealing features, in particular the fact that it is structurally different from the mainstream approach that uses localized qubits.
Indeed in the
mainstream approach, the hardware corresponds directly to each qubit: for instance for spin qubits, one needs a certain number of electrostatic gates per qubit
to confine the electron, address it with microwaves and eventually measure its state.
It follows that the hardware footprint is proportional to the number of qubits.
In contrast, in  this 'synchrotron-like' quantum computer, the flying qubits are stored in a loop  and fast quantum routers are used to bring them to single-qubit gates, two-qubit gates, delay lines or measuring apparatus \cite{takeda_2017}.
Hence, the hardware footprint can in principle be extremely small: a few quantum routers (one per type of gates or measurement) are sufficient to control an arbitrarily large number of qubits.
The Leviton qubits are created on demand and one only needs a loop, which is large enough to hold Levitons while they go around it.

The second, perhaps more important, advantage of this architecture is the connectivity of the two qubit gates: using the delay line shown in the schematic, one could move the qubits so that any pair of flying qubits could be put next to each other and, hence, one could apply two qubit gates between any 2 qubits.
This is again in contrast to the mainstream approach where each qubit being localized, it can only interact with a few other qubits, usually its nearest neighbours.
Such a dramatic increase in connectivity could have deep consequences to reduce the overhead of quantum error correction and fault tolerant operations.

Another advantage of the flying qubit architecture for quantum computing is that
qubits can easily be recycled: old qubits can be expelled from the loop and fresh
% one
\he{ones}
incorporated while ancilla ones can be used to calibrate or test the various parts of the circuit in order to isolate
\he{and retune sections}
% parts
that are not performing correctly.
% or retune them to a correct regime.
This flexibility could again be very instrumental in quantum error correction in order to get rid of rare lethal errors.
Indeed, in quantum error correction, not all errors are equal; some, even if rare, are lethal to the calculation \cite{Waintal_2019}.
In this respect, a long-term advantage of the flying qubit architecture is the possibility to correct these rare errors. Altogether a functional flying qubit technology could make quantum error correction affordable, bringing the millions of qubits which are required to build a fault tolerant quantum computer down to tens of thousands.
Alternatively, the electron-flying-qubit approach could be used to complement other approaches by providing a `quantum bus' that implements the missing long-range coupling between distant localised qubits.
Experimentally, we are still in the early stage of the development of such a \he{electron-flying-qubit}
% flying qubit
platform.
Yet, it is very interesting and appealing to see that it leads to a conceptually very distinct object from the localized qubit approach.
This means in return that there is a lot of room for \he{a} new architecture to be invented to bypass the intrinsic limitations of the ones that are pursued so far.

\he{
	To end this section on the experimental progress on electron flying qubits realised in semiconductor devices, we would also like to point out promising approaches to manipulating single electrons on other unique platforms.
	Alternative to the here-described semiconductor devices, single electrons can be confined on the surface of liquid helium
	%Precise Measurement of Effective Mass of Positive and Negative Charge Carriers in Liquid Helium II
	%J. Poitrenaud and F. I. B. Williams
	%Phys. Rev. Lett. 29, 1230 – Published 30 October 1972
	%
	%Williams_1985
	%Dynamical Hall effect in a two-dimensional classical plasma
	%DC Glattli, EY Andrei, G Deville, J Poitrenaud, FIB Williams
	%Physical review letters 54 (15), 1710 (1985)
	%
	%D. B. Mast, A. J. Dahm, A. L. Fetter,
	%Phys. Rev. Lett. 54, 1710 (1985);
	%
	\cite{Glattli_1985, Mast_1985, Dahm_1987, Byeon_2021} or rare earth atoms such as neon, argon or krypton \cite{Zhou_2022}.
	These systems provide a two-dimensional electron system with ultra-high mobility and strong Coulomb interaction.
	Similar to SAW-driven single-electron transport discussed in section \ref{sec:SAW}, electrons on the surface of liquid helium can be transported with very high precision through coupling to an evanescent piezoelectric SAW \cite{Byeon_2021}.
	Besides that, electrons can be attracted to the surface of a solid crystal made from rare-earth atoms in vacuum.
	For the case of a solid neon substrate, a single electron has been trapped with electrostatic gates and coupled to a superconducting microwave resonator \cite{Zhou_2022}.
	This allowed to observe coherent coupling of motional electron states to a single microwave photon with coherence properties similar to state-of-the-art charge qubits \cite{Chatterjee_2021}.
}
% \cb{
% To end this section on electron charge qubits, let us point out other very promising approaches to manipulate single electrons on other unique platforms.
% Single electrons can actually be confined on the surface of liquid helium \cite{tito-williams, Byeon_2021}
% }
% \cbc{Christian: can you add the reference of seminal work on that: Toto Williams ?}
%
%physics Today 40, 2, 43 (1987); https://doi.org/10.1063/1.881098
%
%
% or rare earth atoms such as neon, argon or ktypton \cite{jin_2022}. }
%

% \cb{
% These systems form an ultra-high mobility, two-dimensional electron system with strong Coulomb interaction.
% Similar to the approach discussed previously,  electrons on the surface of liquid helium can be transported with very high precision through coupling to an evanescent piezoelectric SAW \cite{pollanen_2021}.
% An other experiment has exploited the fact that electrons can be attracted to the surface of a solid crystal made from rare-earth atoms in vacuum.
% For the case of a solid neon substrate a single electron has been trapped with electrostatic gate electrodes and coupled to a superconducting microwave resonator \cite{jin_2022}.
% This allowed to observe coherent coupling of motional electron states to a single microwave photon with coherence properties similar to state of the art charge qubits \cite{Kuemmeth_2021}.
% }
%
%Semiconductor qubits in practice
%Anasua Chatterjee, Paul Stevenson, Silvano De Franceschi, Andrea Morello, Nathalie P. de Leon & Ferdinand Kuemmeth
%
%Nature Reviews Physics volume 3, pages 157–177 (2021)Cite this article
%

\section{Numerical modelling of quantum devices}

\begin{figure}[!h]
	\centering
	\includegraphics[width=11cm]{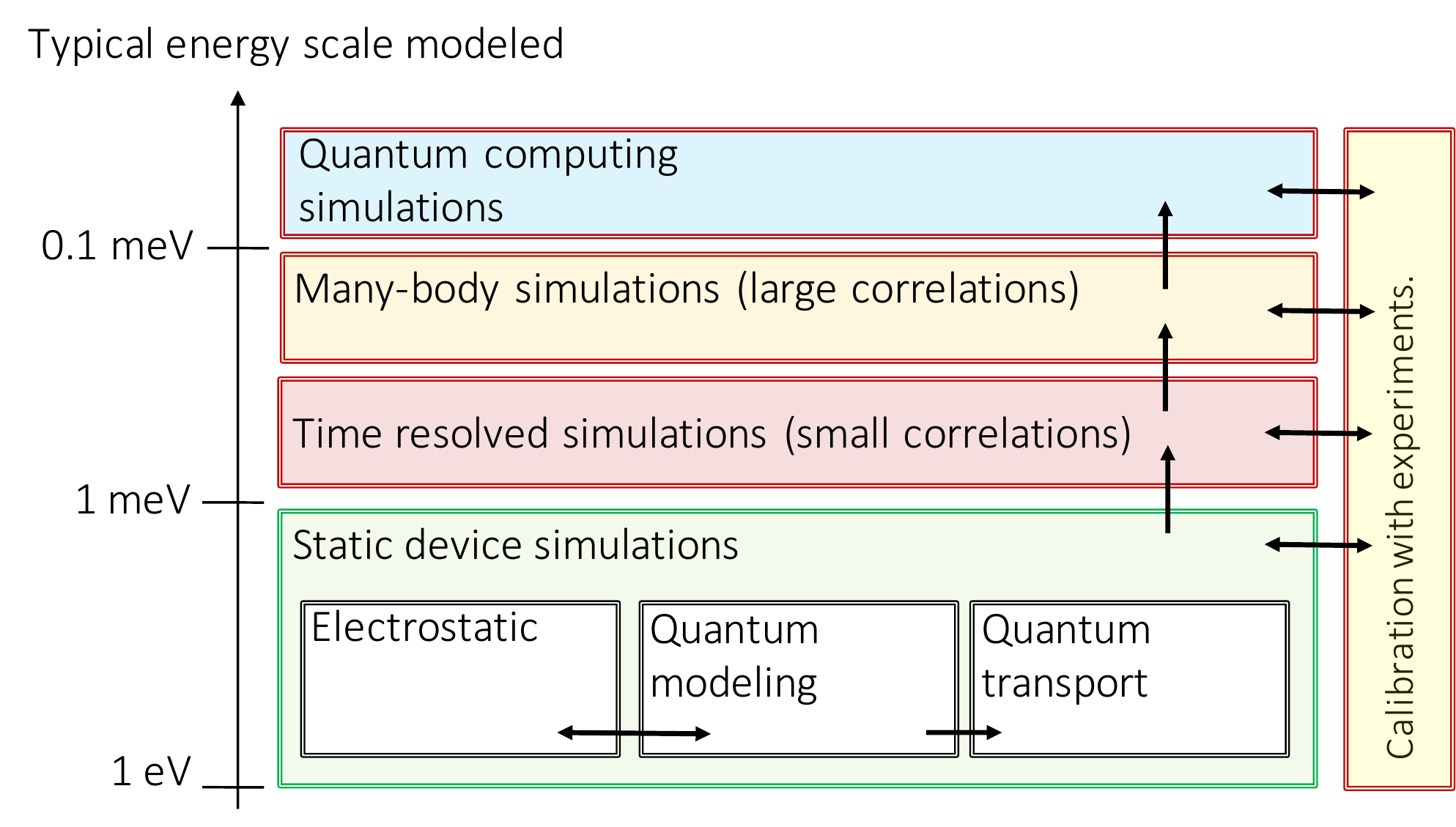}
	\caption{
		\textsf{\textbf{Schematic of a full simulation stack for flying qubits architectures}}
	}
	\label{fig:softwarestack}
\end{figure}

Numerical simulations play an important role in the development of quantum computing architectures and the flying qubit platform is no exception.
Achieving a full stack of numerical tools to compute and predict the various properties of the devices is key to certify that the devices behave as they are supposed to and allows one to eventually optimise their behaviour.
Figure~\ref{fig:softwarestack} shows a typical stack that is being developed for flying qubit architectures.
At the bottom are the device simulations that incorporate the material modelling as well as the geometry of the device.
These are
% \sout{typically}
usually performed at the self-consistent electrostatic quantum level, i.e. the electrostatic problem is solved simultaneously with the quantum problem associated with the active part of the device (typically the region around the GaAs/AlGaAs interface in the devices discussed in this
article).
The self-consistent potential can be used by quantum solvers to calculate the quantum transport properties of the device, e.g. the conductance or the current noise or other observables.
Those properties can be directly compared to DC experimental measurements to obtain a direct feedback on the quality of the modelling and its calibration.
The proprietary {\it nextnano}~\cite{Trellakis2006} platform or the open-source KWANT software \cite{groth_2014} are complementary tools that can be used for this stage.

Once the static properties are well understood, one can proceed to simulate the propagation of the electron flying qubits, including voltage pulses and the associated Levitons, in real time. The TKWANT extension \cite{kloss_2021} of KWANT provides the necessary environment for such simulations (e.g., to study the role of Coulomb repulsion at the time-dependent mean field level).
The next level is a proper treatment of many-body effects aiming to account for e.g. interactions between different Levitons or various
% \sout{relaxation/dephasing}
relaxation and dephasing mechanisms (such as one electron decaying into two electrons and one hole).
We note that there are no general purpose simulation approaches that can handle this problem in a ``blackbox'' way.
At the top of the stack are ``pure'' quantum computer simulators where the actual underlying physics has been hidden and one simulates only the effective dynamics of the computational degrees of freedom (potentially with some extra noise or dissipation terms to account for the actual limitations of the devices).
As indicated by the arrows on the schematic, the different parts of the stack provide parameters to calibrate the other levels.
As one goes up the stack, one usually must give up some microscopic details in order for the computations to remain affordable.
Therefore, the calibrations must be done with care for the errors not to accumulate.
Below, we focus at first on the static simulation part of the stack with a special emphasis on the calibration of the simulations with respect to the experiments and on the modelling of real nanodevices.
At the end of this section, we briefly address time-resolved and many-body simulations.

\subsection{Static quantum mechanical simulations}

\he{
	Tuning a single qubit into optimal operation is so far a tedious task.
	An attempt to find such conditions trying various setups at random is time
	% - \SB{Do we need the dash after time?} % H.E.: No.
	and resource consuming.
	In order to go easier beyond experimental proof of concept (also known as Technology Readiness Level (TRL) 3), it is thus crucial being able to predict the viability of a certain sample design prior to its physical realisation.
	As outlined above, precise potential calculations combined with dynamic quantum mechanical simulations are playing a key role in this regard enabling validation of electron-flying-qubit technology in the lab (TRL 4).
	Being able to predict the reliability of a certain sample geometry paves the way to implement and setup electron flying qubits in a reproducible manner -- enabling validation of the technology in a demonstrative or even commercial setting (TRL 5 and 6).
}

Since the basic elements of the electron flying qubits
(interconnects, TCWs and interferometers) are exploiting
to a great extent
\he{single-particle physics,}
% \sout{the single-particle physics,}
they require high-quality quantum mechanical simulations for one electron in complicated electrostatic potentials.
The necessary information is obtained from a numerical solution of the stationary Schr{\"o}dinger equation, see Eq.~(\ref{Schr_Eq}) below.
The precise solution can be found by using a platform such as the nextnano software.
Its advantages include
% \sout{a}
\he{the} possibility to adapt the numerical procedure for different materials, various geometries of the nanoconductors and shapes of the gate-induced potentials.

Let us review \he{the} basics of the static quantum mechanical simulations, some features of the nextnano tools, and provide examples of how these tools can be used for calibration of the experiments and engineering the nanodevices.

\subsubsection{\he{Basic equations and methods of static single-particle simulations}}

Basic targets of the static quantum mechanical simulations include the study of the shape of electron wave functions and the energy-dependent transmission through one nanounit and through the entire circuit~\cite{BautzePRB2014,WestonWaintal2016,RossignolPRB2018}.
\he{The transverse profile (i.e. along the direction being perpendicular
	to the propagation direction)}
allows one to judge whether the quantum wires are close to the desirable
%%%
% (1D)
%%%
setup and to control, e.g., the absence (presence) of tunnelling between
two isolated (coupled) wires.
Ballistic transmission of the electron through the circuit is even more important.
When various units are connected, there are always spatial inhomogeneities which \he{can} result in \he{reflection}
% \sout{the reflection}
of the propagating electron. % H.E.: It depends: For SAW transport, for instance, potential inhomogeneities do not harm the propagation of the quantum state if they are small enough -- meaning that adiabatic transport is possible. For Levitons it's a different story.
The reflection hinders the flying qubits from their normal operation and must be minimised as much as possible. % H.E.: This is true! But a bit simplified, since every transport approach has its particularities ... we can keep it ...
To this end, one can numerically find an energy corresponding to transparency windows for a realistic circuit and to work further in a vicinity of this special energy \he{range}. %H.E.: I think there is some operational range than a specific value ...

\he{The quantum mechanical system}
% \sout{The quantum mechanics}
can be modeled at different levels of approximations that range from a semi-classical description to an effective mass approximation to a multi-band $k \cdot p$ model.
Considering conduction-band electrons within the single-band approximation, the envelope wave functions, $ \psi_n $, are solutions of the stationary Schr{\"o}dinger equation
\he{
	\begin{equation}
	\label{Schr_Eq}
	\hat{H} \psi_n(\mathbf{x}) = E_n \psi_n(\mathbf{x}) ,
	\end{equation}
	where $\hat{H}$ is the Hamiltonian operator of the closed quantum system, $ E_n $ are the energy levels defining the energy spectrum of the system, $ n $ are quantum numbers marking different single electron quantum states, and $ \mathbf{x} = ( x, y, z ) $ is the space coordinate.
	The Hamiltonian operator
	\begin{equation}
	\label{Schr_EQ_Sum}
	\hat{H} = \epsilon(\hat{\mathbf{p}}) + V(\mathbf{x})
	\end{equation}
	is the sum of a kinetic energy operator $\epsilon(\hat{\mathbf{p}})$
	and a potential energy $V(\mathbf{x})$, where the electron momentum operator is defined in the standard way as $ \hat{\mathbf{p}} = - i \hbar \mathbf{\nabla}$.
	Here, $\epsilon(\hat{\mathbf{p}},\mathbf{x})$ is the (possibly position-dependent) dispersion relation describing the momentum dependence of the electron energy which accounts for all effects governed by the crystalline lattice, and $ V(\mathbf{x})$ is the inhomogeneous potential in which the electron propagates.
	$ V(\mathbf{x})$ contains the electrostatic potential and conduction band offsets at material interfaces.
	For example, in the simple case of a homogeneous isotropic material where the electrons move almost freely, one can use the effective mass approximation which yields
	$ \epsilon(\mathbf{p}) = (p_x^2 + p_y^2 + p_z^2) / 2 m^* $, where $ m^*$
	is the effective mass of the electron in the material.
	
	The potential $\phi(\mathbf{x})$ describes the electrostatics within
	the system and is the solution of the Poisson equation
	\begin{equation}
	\label{Poisson}
	\nabla \cdot \left[ \varepsilon (\mathbf{x}) \nabla \phi(\mathbf{x})  \right] = -\rho(\mathbf{x}),
	\end{equation}
	where $\varepsilon(\mathbf{x})$ is the material-dependent permittivity, and $\rho (\mathbf{x})$ is the charge density throughout the system. This charge density is given by
	\begin{equation}
	\label{ChargeDensity}
	\rho ({\mathbf{x}}) = e \left[ -n (\mathbf{x}) + p (\mathbf{x}) + N_{{\text{D}}}^{+}(\mathbf{x}) - N_{{\text{A}}}^{-} (\mathbf{x})  \right]
	+ \rho_{\textrm{fixed}} ({\mathbf{x}}),
	\end{equation}
	where $n(\mathbf{x})$ and $p(\mathbf{x})$ are the electron and hole densities, and $N_{{\text{D}}}^{+}(\mathbf{x})$ and $N_{{\text{A}}}^{-}(\mathbf{x})$ are the ionized donor and acceptor concentrations, respectively, $e$ is the (positive) elementary charge, and $\rho_{\textrm{fixed}} ({\mathbf{x}})$ contains immobile space or surface charges.
	
	Here, the electron density $n(\mathbf{x})$ explicitly depends on the energy levels $E_n$ and envelope wave functions $\psi_n$ from Schr{\"o}dinger's equation, Eqs.~(\ref{Schr_Eq},\ref{Schr_EQ_Sum}).
	For a finite system at equilibrium, the electron density is given by
	\begin{equation}
	\label{qmdensity}
	n (\mathbf{x}) = \sum_n  \frac{2 |\psi_n(\mathbf{x})|^2}{1 + \exp((E_n - E_\textrm{F})/k_\textrm{B}T)} ,
	\end{equation}
	where $E_\textrm{F}$ is the Fermi level (or chemical potential), $T$ is the temperature, and $k_\textrm{B}$ is the Boltzmann constant.
	Thus, the electrostatic potential $\phi(\mathbf{x})$ depends
	on the energy levels $E_n$ and wave functions $\psi_n$, but also enters the Schr{\"o}dinger equation, Eqs.~(\ref{Schr_Eq},\ref{Schr_EQ_Sum}),
	as part of the potential energy operator $V(\mathbf{x})$. This shows that the Schr{\"o}dinger equation
	and Poisson equation (Eq.~(\ref{Poisson})) are coupled and need to be solved self-consistently.

	The self-consistently obtained spectrum and wave functions can be used further to calculate quantities which explain and describe quantum transport through various nanodevices connected to external leads.
	Two such quantities are (i)~the partial local density of states (pLDoS), $ n(x,y, E)$, and (ii)~the energy dependent transmission, $ T_{ij}(E) $.
	The {\it pLDoS} is the probability to find in a given space-point the propagating electron which has a given energy, $E$, and has been injected in a given lead.
	We note in passing that the local density of states is the sum of the pLDoS over all leads.
	Hence, the coordinate dependence of pLDoS illustrates how the electron with a given energy propagates through the device \cite{datta_2005}.
	{\it The energy dependent transmission}, $ T_{ij}(E) $, is determined by the probability for the electron which is injected into lead $ i $ to reach lead $ j $.
	
	The pLDoS and the transmission from one lead to another can be found by using
	the retarded
	Green's function, $ \hat{G}^R(E) = ( [E + i \alpha] \hat{1} - \hat{H} )^{-1} $ where
	$ E $ is the electron energy and
	$\alpha \to 0^{+}$
	% $ \alpha \to + 0 $
	is a mathematical regularizer
	which reflects the retardation of the physical response (see Ref.~\cite{datta_2005}
	and Ref.~\cite{birner_2009} for details). In the space-coordinate representation,
	the coordinate-dependent Green's function can be expressed via the wave functions
	and the spectrum (the so-called spectral and Lehmann representations).
	Hence, the solution of the Schr{\"o}dinger equation provides
	the input needed for the theoretical study of transmission.
	
	The transmission provides valuable information on quantum interference occurring
	in the TCW or the Aharonov--Bohm (AB) interferometer. In the original setup,
	the AB interferometer involves the magnetic field,
	$ \mathbf{B} = {\rm curl} \mathbf{A} $, which
	can be included into the study as a shift of the momentum operator by the
	vector potential, $ \hat{\mathbf{p}} \to \hat{\mathbf{p}} - e \mathbf{A} $.
	% \oy{\sout{, with $ e $ being the electron charge.}}
	% vector potential, $ \hat{\mathbf{p}} \to \hat{\mathbf{p}} - e / c \mathbf{A} $, with
	% $ e $ and $ c $ being the electron charge and the speed of light.
	In practice, a magnetic field variation is too slow on the time scales needed
	for qubit operation, so electrostatic manipulation of the gates is much more practical.
	Nevertheless, for optimisation of the design, experiments and simulations in the presence of a magnetic field are still useful.

	Schr{\"o}dinger's equation can only be solved analytically for some specially chosen potentials, whereas,
	in the general case, spectra and wave functions can only be found numerically.
	The numerical solvers are applied after discretization, which means that the continuous
	space is reduced to points on a grid and derivatives are substituted by differences.
	Here, the grid spacing is an important parameter which controls the accuracy of the numerically obtained answers.
	In our qubit devices, layer structures and dopant distributions create a triangular shaped quantum well along the substrate growth direction.
	In this quantum well, quantum confinement effects cause the electrons to form a two-dimensional electron gas
	which is modulated in the two directions perpendicular to the substrate growth direction in accordance with the influence of the gate geometries.
	For such 3D devices where thousands of eigenstates have to be taken into account, efficient solvers for the Poisson and Schr{\"o}dinger equations
	such as preconditioned conjugate gradient for Poisson and Arnoldi iteration for Schr{\"o}dinger are mandatory in order
	to overcome the huge computational costs.
	Moreover, achieving self-consistency between the Poisson and the Schr{\"o}dinger equation is not easy and requires
	the use of special techniques such as predictor--corrector methods~\cite{trellakis_iteration_1997} in order to robustly obtain solutions. In strongly non-linear regimes such as in the quantum Hall regime,
	other techniques such as \cite{Armagnat_2019} might be needed.
	
	As we have already mentioned, the
	simulation of quantum transport and thus obtaining the pLDoS and the transmission requires the use of Green's function techniques \cite{GrangeAPL2019}, which are computationally extremely expensive in the most general case. Fortunately, the ballistic limit of quantum transport suffices for the accurate description of flying qubits. This allows the so-called Contact Block Reduction (CBR) method~\cite{MamaluyJAP2003,birner_2009} to be used here in order to reduce the computational cost down to a point that even large three-dimensional devices of arbitrary shape and with an arbitrary number of contacts can be easily modelled.
}

\subsubsection{The nextnano software and its applications for engineering flying qubits}
% \sout{The nextnano software is developed by the company nextnano GmbH and is a user-oriented platform meant for modelling }
\he{Starting from the year 2000, the nextnano software had been developed at the Walter Schottky Institute of the Technische Universit\"at M\"unchen.  Later, it resulted in the spin-off company nextnano GmbH.
	The software, which is now further developed by this company,}
% OYe and SB prefer previous version
%\he{The nextnano software initially emerged from the Walter Schottky Institute of the Technical University of Munich in 2000, from which the spin-off company nextnano GmbH emerged, which is now developing the software further.}
is a user-oriented platform meant for modelling
various semiconductor-based nanodevices, cf. Refs.~\cite{birner_2006,birner_2007,Trellakis2006},
including optoelectronic elements and qubits.
The main focus is on the simulation
of the quantum mechanical properties of such devices. The nextnano's core product,
the nextnano++ software~\cite{Trellakis2006}, is a 3D Schr\"odinger--Poisson--Current\he{/CBR}
solver for nanotransistors, LEDs, laser diodes, photodetectors, quantum dots, nanowires,
solar cells and qubits.
The second product, nextnano.NEGF~\cite{GrangeAPL2019},
is a quantum transport solver targeting quantum cascade lasers and resonant tunneling diodes.
The nextnano software (including its early versions) has been successfully used to optimise
the design of semiconductor-based (charge and spin) qubits
\cite{ZiboldPRB2007,CaflischSIAM2005,Wild_2010,Ramirez_2019,BuonacorsiPRB2020, jirovec2021singlet}. Below, we
focus on several important applications of this software for engineering the electron flying qubits.

% \OYe{@Stefan: The above text on nn could be extended / improved}

\textbf{Appropriate models for quantitative simulations}:
Quantum devices made from GaAs semiconductor heterostructures can easily be engineered by proper design of the gate geometry.
To ensure the best performance of the electronic device -- that is to find the most suitable gate geometry -- it is crucial to know the exact electrostatic potential landscape generated by the electrostatic gates.
This requires  to take into account the material parameters such as 2DEG density and mobility, dopants concentration, induced surface charges, etc.

\begin{figure}[!h]
	\centering
	\includegraphics[width=7cm]{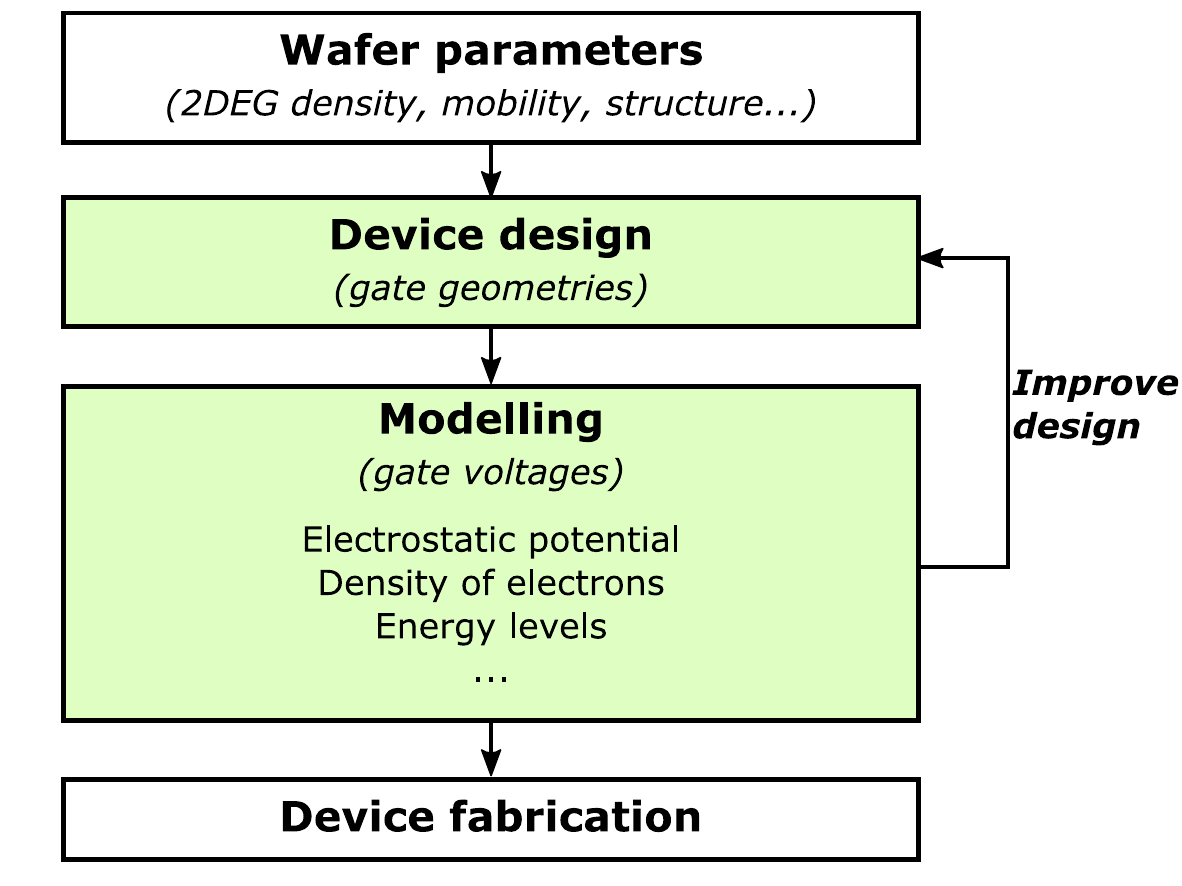}
	\caption{
		\textsf{\textbf{Schematic of the ideal workflow from the sample design to the device fabrication}}
	}\label{fig:nn_workflow}
\end{figure}

Traditionally, the workflow to determine the optimum gate geometries for a given heterostructure has been an iterative process between device fabrication in clean room facilities and low-temperature characterisations. This is immensely time consuming and resource demanding.
The ideal workflow is \he{presented} in Fig.~\ref{fig:nn_workflow} where the iterative process takes place mainly at the modelling stage, \textit{before} the device fabrication.

\begin{figure}[!h]
	\centering
	\includegraphics[width=11cm]{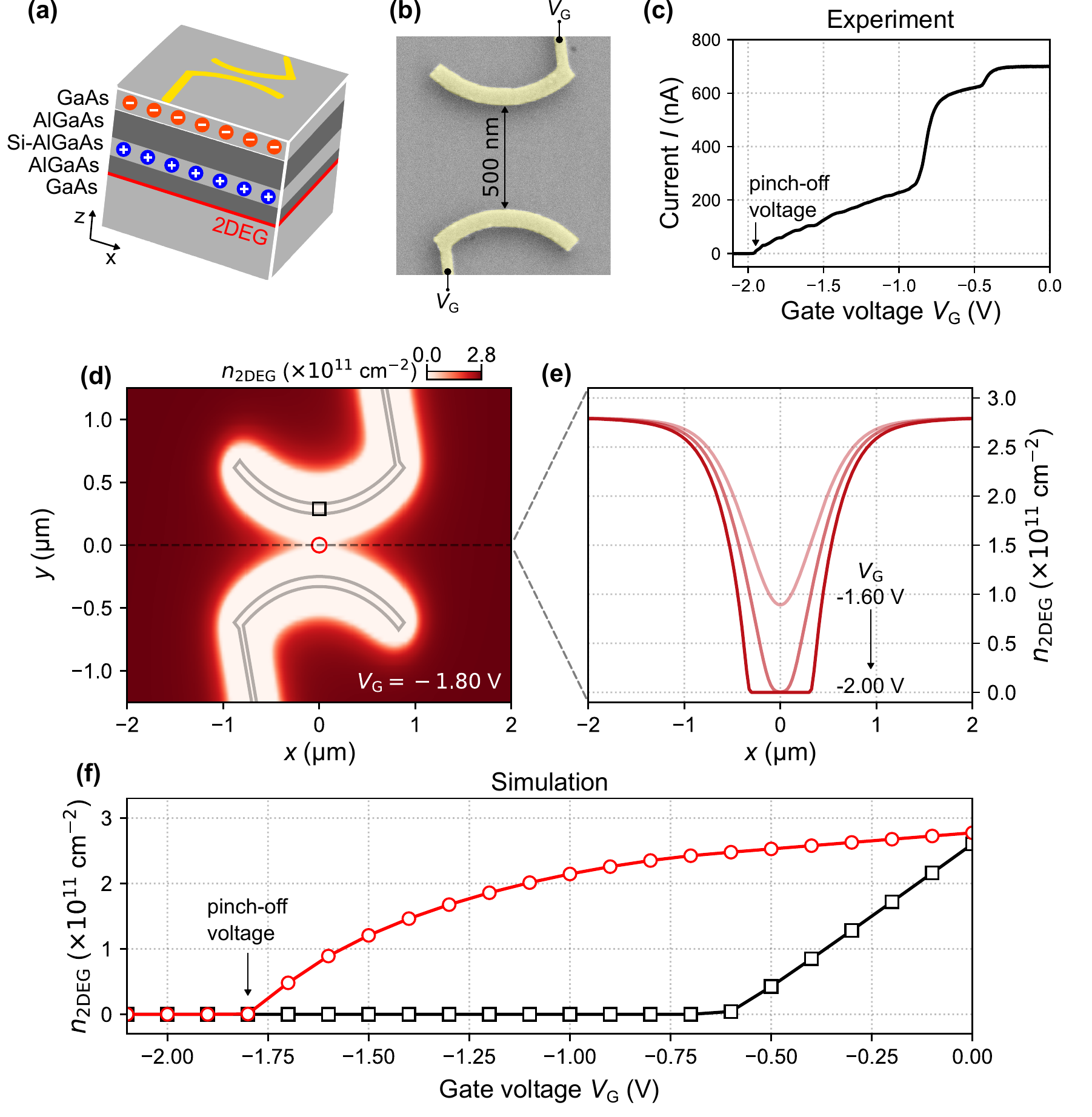}
	\caption{
		\textsf{\textbf{Calibration details.}
			\textsf{\textbf{(a)}}
			Schematic of the GaAs/AlGaAs heterostructure below the surface-gate defined QPC (yellow) with indications on the surface charges (red circles), Si dopants (blue circles) and the 2DEG (red line).
			\textsf{\textbf{(b)}}
			Scanning-electron-microscopy image of a QPC design.
			\textsf{\textbf{(c)}}
			Experimental current $I$ across the narrow constriction as a function of the surface-gate voltage $V_{\rm G}$.
			\textsf{\textbf{(d)}}
			Simulated distribution of the electron density $n_{\rm 2DEG}$ for $V_{\rm G}=-1.80$ V using nextnano. The grey polygons correspond to the gate geometry.
			\textsf{\textbf{(e)}}
			\he{Electron density} $n_{\rm 2DEG}$ along the constriction ($y=0$; dashed line in (d)
			% \SB{Comment: I think it must be (d) and not (a) here.})
			for three values of $V_{\rm G}$.
			\textsf{\textbf{(f)}} Electron density $n_{\rm 2DEG}$ below the surface gate (black square) and at the middle of the constriction (red circle). The simulated QPC pinch-off occurs when the 2DEG is completely depleted. }
	}
	\label{simulations}
\end{figure}

To find an accurate model for quantitative simulations, Chatzikyriakou \emph{et al.} \cite{Chatzikyriakou_2021} developed a model using the nextnano software and benchmarked
it with experimentally measured QPCs with a wide range of geometries.
They
% assume
\he{assumed}
a layer of surface charges and a spatially uniform doping concentration, both having a frozen ionization state due to the very low temperatures at which the experimental measurements are taken \cite{takada_2019,hou_2018}.
First, 1D simulations of GaAs/AlGaAs heterostructures (Fig.~\ref{simulations}a), with a Schottky gate on top, are employed in order to deduce the doping concentration such that the simulation reproduces experimentally measured characteristics of these heterostructures that exist on the chip that hosts the QPCs. These structures are covered by metal electrodes that are very large compared to the QPC gates ($>$ 500 nm in each Cartesian direction) and that are finally connected to the QPC gates. Then, removing the gate from the simulated heterostructure, the surface charges are adjusted so that the 2DEG electron density is equal to that taken from Hall measurements on the same wafer, at \he{$ T = 4.2$~{\rm K}} (frozen surface states).
To simulate the region where electron transport takes place, 3D simulations are carried out with the exact gate geometry of the quantum device. These gate geometries are directly imported from the
\he{computer-aided design (CAD) layouts}
% \sout{CAD files}
(standard files in GDS format) using the open-source Python package nextnanopy~\cite{nn_nextnanopy}.

Figure~\ref{simulations} shows a typical example of electron depletion in the 2DEG when applying a gate voltage to the electrostatic gates that face each other.
When a small negative voltage is applied, the electron density under the gates is first depleted, forming only a narrow 1D constriction in the center of the two gates.
Reducing further the voltage, the 1D channel is completely depleted and the transport channel is \textit{pinched-off}.
The simulation shows a remarkable agreement with the experimental pinch-off value \he{($ V_{\rm{G}} = -1.8 \rm{~V}$)}.
% ($ V_{\rm{G}} = -1.25 \rm{~V}$).

\begin{figure}[!h]
	\centering
	\includegraphics[width=11cm]{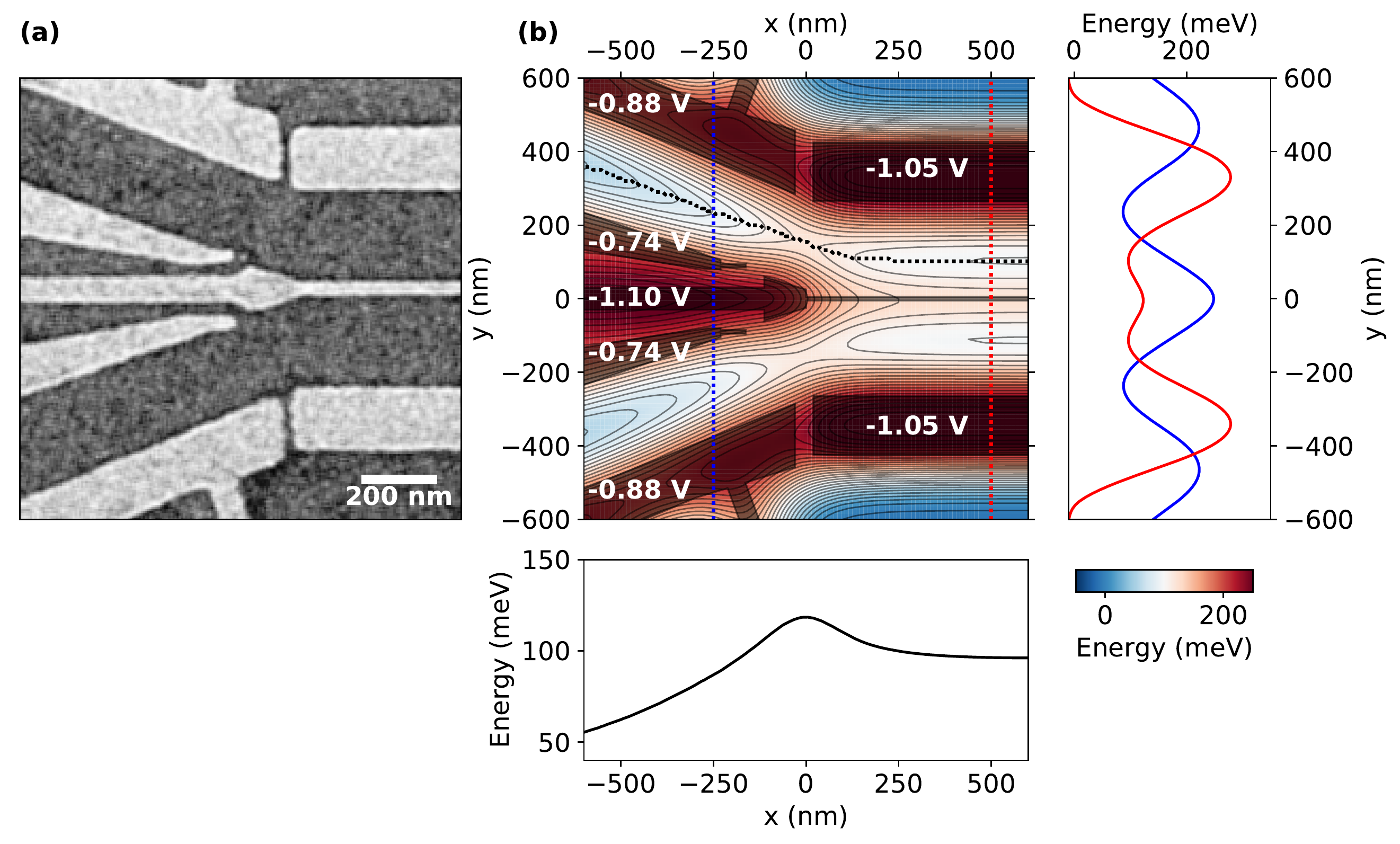}
	\caption{
		\textsf{\textbf{Electrostatic potential landscape.}}
		\textsf{\textbf{(a)}} Scanning electron micrograph of the entrance of a tunnel-coupled region. Adapted from Ref.~\cite{takada_2019}.
		\textsf{\textbf{(b)}}
		Electrostatic potential induced at the 2DEG from 3D simulations using realistic gate geometries (fainted black layout) and typical voltage values applied to the electrostatic gates. Equipotential lines are shown as continuous lines. Vertical cuts (blue and red) show the double-well potential before and within the tunnel-coupled region. The black line represents the path which follows the minimum in the potential landscape.
	}
	\label{fig:jw_sim}
\end{figure}

In the same spirit, one can use such simulations to calculate the potential variations seen by the electrons within the 2DEG.
An example of a complex quantum device with a tunnel-coupled wire from Takada \emph{et al.} \cite{takada_2019} is shown in Fig.~\ref{fig:jw_sim}.
Using the exact gate geometries and voltages from the experiment, the electrostatic simulations reveal the variation of the potential along the path which an electron would follow before entering the tunnel-coupled region (black line).
In these experiments, the single electron is excited to higher energy states which were attributed to the abruptness in the potential landscape at this location.
This undesired excitation could be mitigated by optimising the device geometries thanks to quantitative modellings.

\begin{figure}[!h]
	\centering
	\includegraphics[width=9cm]{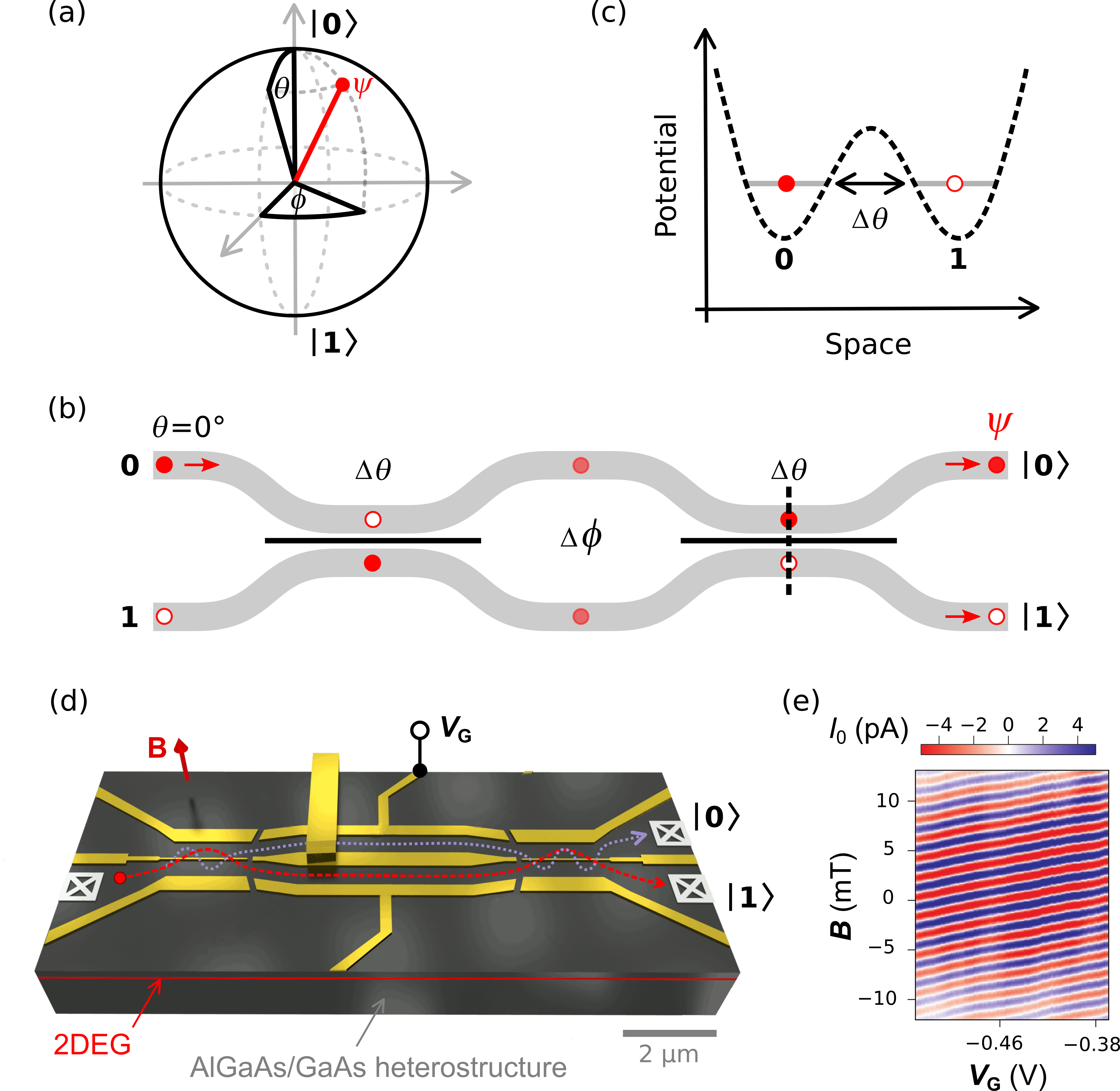}
	\caption{
		\textsf{\textbf{Electron flying qubit.}}
		\textsf{\textbf{(a)}}
		Bloch sphere showing the qubit state $\psi$ that is defined via the tunnel-coupling $\theta$ and the phase shift $\phi$.
		\textsf{\textbf{(b)}}
		% \SB{Comment: We need ket symbols for 0 and 1 on the left injection leads, similar as on the right side; Insert ket symbols also on Fig. c)}
		% \comment{H.E.: Why do we need that? The idea here is that we start from a classical state (without kets) and make it quantum (with kets) by passing it through the interferometer. Compare potential in Fig. 1c.}
		Schematic of an electron two-path interferometer allowing precise adjustment of the qubit state via tunnel-coupling ($\Delta\theta$) and the quantum phase ($\Delta\phi$) picked up along transport across the central island.
		\textsf{\textbf{(c)}}
		Transverse potential landscape across the tunnel-coupling region with schematic indications of the qubit basis 0 and 1.
		\textsf{\textbf{(d)}}
		Detailed 3D geometry of a quantum interferometer showing surface-gates (golden) defining the potential landscape and thus the transport paths in the two-dimensional electron gas (2DEG) located below the surface.
		The white crossed boxes indicate ohmic contacts enabling electrical connection to the 2DEG.
		\textsf{\textbf{(e)}}
		Measurement of quantum oscillations as function of a perpendicular magnetic field $B$ and the side-gate voltage $V_\textrm{G}$.
		Data adapted from Ref.~\cite{yamamoto_2012}.
	}\label{fig:bloch}
\end{figure}

\begin{figure}[!t]
	\begin{center}
		\includegraphics[width=12.25cm]{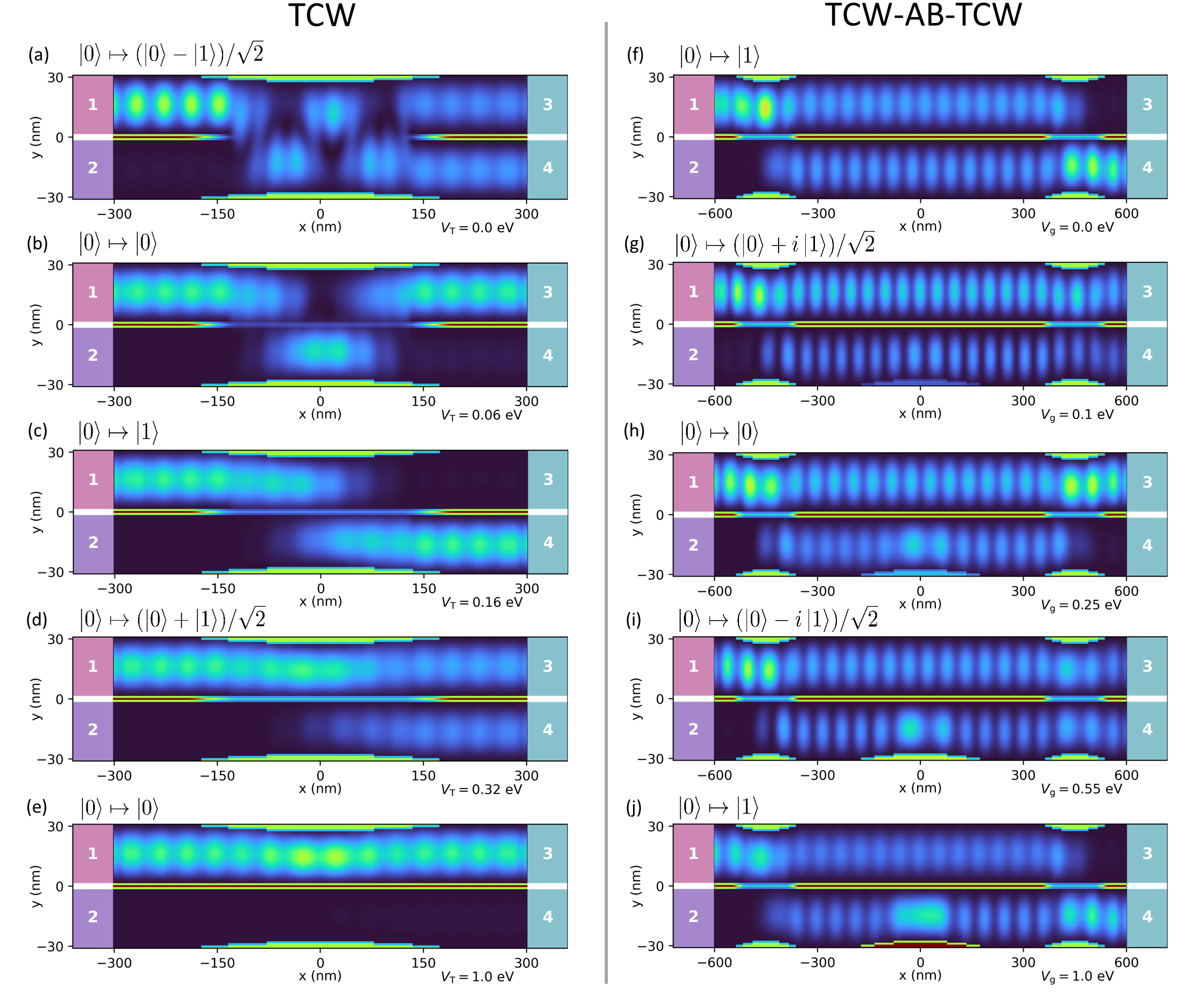}
	\end{center}
	%%%
	%    \includegraphics[width=7cm]{fig12_BS_LDOS_Vt-0.0.pdf}
	%    \includegraphics[width=7cm]{fig12_BS_LDOS_Vt-0.1.pdf}
	%    \includegraphics[width=7cm]{fig12_BS_LDOS_Vt-0.2.pdf}
	%    \includegraphics[width=7cm]{fig12_BS_LDOS_Vt-10.0.pdf}
	%%%
	\caption{
		%%%
		% \he{
		% \textsf{\textbf{nextnano simulation of transmission states.}}
		% Electron partial local density of states (pLDoS), $ n(x,y,E)$, for
		% \textsf{\textbf{(a-e)}}
		% a tunnel-coupled wire (TCW) with increasing barrier voltage $V_{\rm T}$ and
		% \textsf{\textbf{(f-j)}}
		% a quantum interferometer (TCW--AB--TCW) with increasing voltage $V_{\rm g}$ on
		% a side gate of the bottom path.
		% The white numbers indicate the four ohmic contacts of the the four-terminal
		% nanoscale devices.
		% The background shows the potential landscape defined by the voltage on the surface
		% gates.
		% }
		%%%
		\he{
			{\bf nextnano simulations of the electron partial local density of states}
			%%%
			% , $ n(x,y,E)$,
			%%%
			in the tunnel-coupled wire (TCW: panels (a-e)) and the
			TCW -- Aharonov--Bohm interferometer -- TCW nanodevice (TCW--AB--TCW: panels
			(f-j)). Both devices are connected to four terminals (marked by white numbers).
			The background shows the potential landscape defined by the voltage on the
			surface gates, cf. Fig.~\ref{fig:TransmSimulations}(a,d).
			%%%
			% The electrostatic potential in both geometries is shown in the upper % panels of Fig.~\ref{fig:TransmSimulations}.
			%%%
			% OYe & SB: we will adjust the color scheme of figs. 12 & 13
			%%%
			% and described in its caption.
			%%%
			The electron with a given energy ($ E = 9.2 $~meV for TCW and
			$ E = 7.5 $~meV for TCW--AB--TCW)
			%%%
			% for the left and right columns, respectively,
			%%%
			is always injected into the upper incoming channel from the 1\textsuperscript{st} lead, $\ket{0}$ state.
			The states at the output leads are indicated at the top of each panel and explained
			in the main text.
			%%%
			% Upon measurement, the qubit is projected onto either  $\ket{0}$ or
			% $\ket{1}$ state with a probability determined by the pLDoS in the output
			% leads no.~3 and 4, respectively.
			%%%
			{\it Panels} (a-e):
			%%%
			% (from top to bottom)
			%%%
			the pLDoS in TCW for increasing the tunneling barrier voltage
			(described by $ V_{\rm T} $).
			%%%
			% , which are indicated on panels (a-e).
			%%%
			% corresponding output states are:
			% (a)~equal superposition of $\ket{0}$ and $\ket{1}$ in the absence of tunnelling
			% barrier;
			% (b)~$\ket{0}$;
			% (c)~$\ket{1}$;
			%%%
			% (the electron switches from the upper path to the lower one);
			%%%
			% (d)~equal superposition of $\ket{0}$ and $\ket{1}$ in the presence of
			% tunnelling barrier;
			% (e)~$\ket{0}$ (the trivial case where the high barrier disconnects the wires).
			%%%
			{\it Panels} (f-j): the pLDoS in TCW--AB--TCW for increasing voltage on a side
			gate of the bottom path (described by $V_{\rm g} $).
			%%%
			% , which are
			% indicated on panels (f-j); corresponding output states are:
			% (f)~$\ket{1}$ (the gate voltage is not
			% applied, the initial state of the electron is rotated in the TCW regions
			% by $ \pi $,
			% i.e. from the north pole of the Bloch sphere to south one);
			% (g) and (i)~equal superposition of $\ket{0}$ and $\ket{1}$;
			% (h)~$\ket{0}$; (j)~$\ket{1}$ (also governed by TCW regions since the lower
			% path of the AB region is almost blocked by the large gate voltage).
			%%%
			% All panels were generated by using the nextnano software.
			%%%
		}
	}\label{fig:density}
\end{figure}

The shape of the electrostatic potential in the 2DEG plane is input to further calculations of the energy-dependent transmission function which corresponds to the probability that an electron is reflected or transmitted along the different paths in the flying qubit structure.
One-dimensional cuts through this potential in the uncoupled wires (blue) and within (red) the tunnel-coupled regions are shown in Fig.~\ref{fig:jw_sim}b.
One can see parabolically shaped double-well confinement potentials, cf.  Fig.~\ref{fig:bloch}c.
Such potentials and the interplay between symmetric and anti-symmetric states with respect to the direction perpendicular to the propagation direction have been analyzed numerically in Ref.~\cite{BautzePRB2014}
where detailed features of the transport measurements such as in-phase
and anti-phase oscillations of the two output currents as well as a smooth phase shift when sweeping a side gate have been reproduced.
By injecting an electron into the upper rail $\ket{0}$, the wave
function will evolve into a superposition of symmetric and anti-symmetric states.
While travelling through the interaction region,
the wave function of the electron will then pick up a phase
and will evolve into a superposition of $\ket{0}$ and $\ket{1}$~\cite{bauerle_2018} \he{as shown in Fig.~\ref{fig:density} by a simulation example}.
%%%
% A simple example of this evolution while the electron is travelling
% is shown in Fig.~\ref{fig:density}.
%%%

% \cb{Fig 13 is mentioned here but then only briefly at the end. Would it make sense to show fig 13 first and move the description of the end to here ?}
% \\

{\bf Simulations of \he{pLDoS and }transmission through nanodevices}:
Eq.~(\ref{RotState}) describes how the
ideal electron flying qubit is expected to operate:
The electron is injected in either the upper or lower incoming channel
(see  Fig.~\ref{fig:bloch}b) and propagates
% reflectionless
\he{without reflection}
through
the quantum device, where the electron state is rotated in the Hilbert space. This rotation
can be illustrated with the help of the Bloch sphere, Fig.~\ref{fig:bloch}a. Angles $ \theta $
and $ \phi $ are generated by the tunneling regions and interferometers, respectively. The output
state is a coherent superposition of the states $ | 0 \rangle $ and $ | 1 \rangle $. It is
controlled by the tunneling barriers and either by the magnetic field or by the asymmetric bias
of the interferometer. Since using the magnetic field is technologically inconvenient, we focus
in this section on the asymmetrically biased interferometers.

There are two points which
have fundamental importance for engineering the electron flying qubits.
Namely, one needs an experimental setup where, on one hand, the reflection is reduced to a minimum and, on the other hand, the sensitivity of the electron state to the respective gate voltages is high.
Let us explain how using the nextnano software helps to find such a setup.

\begin{figure}[!h]
	\begin{center}
		\includegraphics[width=12cm]{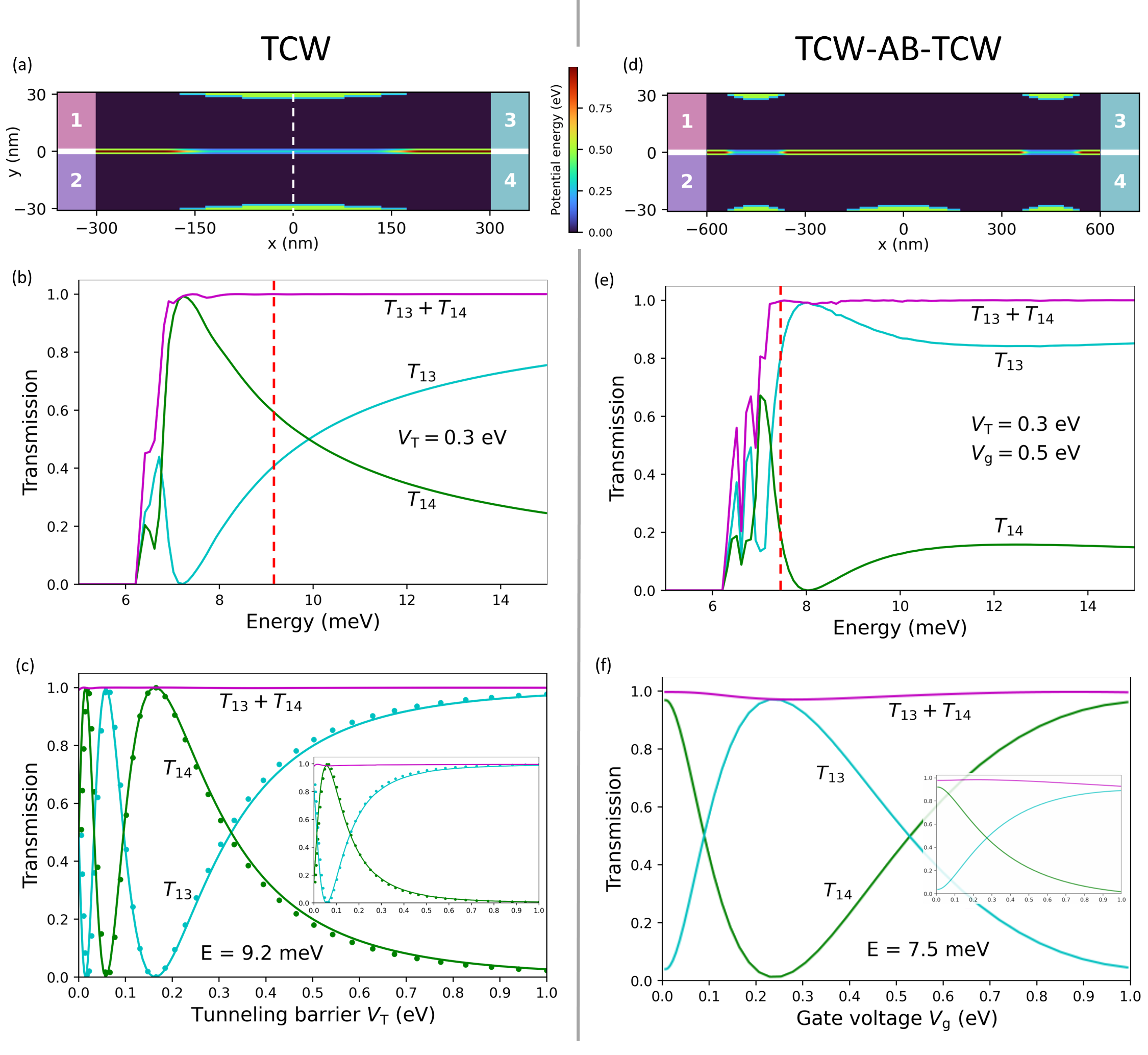}
	\end{center}
	%%%
	%       \includegraphics[width=6cm]{fig13_BS_geom.pdf}
	%       \includegraphics[width=6cm]{fig13_Curc_geom.pdf}
	%    \end{center}
	%    \vspace{0.2cm}
	%    \begin{center}
	%       \includegraphics[width=6cm]{fig13_BS_T_En.pdf}
	%       \includegraphics[width=6cm]{fig13_Curc_T_En.pdf}
	%    \end{center}
	%    \vspace{0.2cm}
	%    \begin{center}
	%       \includegraphics[width=6cm]{fig13_BS_T_Barr.pdf}
	%       \includegraphics[width=6cm]{fig13_Curc_T_Barr.pdf}
	%%%
	\caption{
		%%%	
		%	\he{
		%	\textsf{\textbf{(a,d)}}
		%	Potential landscapes with schematic indications of the four terminals
		%   (white numbers).
		%	\textsf{\textbf{(b,e)}}
		%	Transmission to terminals 3 and 4 as function of the energy of an electron
		%   injected at terminal 1.
		%	\textsf{\textbf{(c,f)}}
		%	Transmission as function of the tunnel barrier $V_{\rm T}$ (TCW) and a
		%   side gate $V_{\rm g}$ (TCW--AB--TCW) at the working points
		%   indicated by the vertical dashed lines in (c,d).\SB{This statement does
		%   not make sense. First, it should probably (b,e) but the vertical lines
		%   have different meaning!}
		%	The insets show simulations of devices with half length.
		% 	The points in (c) show the combination of the semi-phenomenological theory
		%   and the
		%	1D simulation of the spectrum at the center $x=0$ of the device -- see
		%   vertical dashed line in (a).
		%	}
		%%	
		\he{{\bf nextnano simulation of transmission} for the tunnel-coupled
			wire (TCW: panels (a-c)) and the TCW -- Aharonov--Bohm interferometer
			-- TCW nanodevice (TCW--AB--TCW: panels (d-f)). Both devices are connected to four terminals (external leads marked by white numbers).
			%%%
			% TCW contains
			% one region, where two quantum wires are coupled via tunneling,
			% while the interferometer includes two TCW regions embracing a gate-defined
			% counterpart of the Aharonov--Bohm interferometer.
			%%%
			{\it Panels}~(a,d): Potential landscapes of TCW and TCW--AB--TCW devices.
			%%% 	
			% 	with schematic indications of the four terminals (white numbers)
			%%% 	
			% 	An example of the numerical study of electronic transmission
			% 	through two different nanounits connected to four terminals (external
			%	leads): {\it Upper panels} show the geometry of the nanounits, which contain
			% 	either one region, where two quantum wires are coupled via tunneling
			% 	(left upper panel and further the entire left column), or two such regions
			% 	embracing a gate-defined counterpart of the Aharonov--Bohm interferometer
			% 	(right upper panel and further the entire right column).
			% 	Leads are marked by white numbers.
			%%%
			Red and light blue separation regions denote
			impenetrable (very high with the height $ V_\infty = 10\rm{~eV}$)
			and penetrable (tunneling with the height $ V_{\rm{T}} $) potential
			barriers, respectively. Green regions mark those parts
			of the device where the gate voltages $ 0.5\rm{~eV} $ and $ V_{\rm{g}} $ are
			applied. {\it Panels~}(b,e): Energy-dependent transmission
			of the electron from the lead no.~1 into the leads no.~3 ($ T_{13}$) and
			no.~4 ($ T_{14}$).
			%%%
			% The energy of the incoming electron is counted from
			% the potential energy of the lead no.~1.
			%%%
			% All gate voltages are fixed.
			%%%
			Red dashed lines mark
			some electron energies where the reflection is almost absent,
			$ T_{13} + T_{14} \simeq 1 $ ($ E = 9.2 $ meV for TCW and $ 7.5 $ meV for TCW--AB--TCW).
			{\it Panels~}(c,f): Almost reflectionless transmission
			of the electron with fixed energy as a function of
			%%%
			% the tunneling barrier height
			%%%
			$ V_{\rm{T}} $ (TCW) and
			%%%
			% the central gate voltage
			%%%
			$ V_{\rm{g}} $ (TCW--AB--TCW). Dots in panel (c)
			correspond to the semi-phenomenological theory supplied by the
			1D simulation of the spectrum at the center of the device, $ x = 0 $ (marked by
			the dashed line in panel (a)).
			{\it Insets}: The same dependence as in the main figures but for devices
			with half length, where the accessible number of quantum oscillations is
			much smaller.
			%%% 	
			% 	which have either half the total length in $x$ direction
			% 	(TCW) or half the length only in the AB region (right panel).
			%%%
			% 	More detailed description of the numerical simulations can be found
			%	in the main text.
			%%%
		}
	}
	\label{fig:TransmSimulations}
\end{figure}

\he{
	To this end, nextnano enables the calculation of the pLDoS and the transmission of the nanodevices using the CBR method \cite{birner_2009,MamaluyJAP2003}.
	In the following, we demonstrate such calculation via two cases that are the central building blocks of the electron-flying-qubit architecture.
	Firstly (see left columns of Fig.~\ref{fig:density} and
	Fig.~\ref{fig:TransmSimulations}), we address electron propagation through a tunnel-coupled wire (TCW).
	We remind readers that here at $y=0$ a narrow potential barrier is present between two transport channels that allows for coherent tunneling of the electron.
	Secondly (see right columns of Fig.~\ref{fig:density} and Fig.~\ref{fig:TransmSimulations}), we investigate electron propagation through a quantum interferometer (compare Fig.~\ref{fig:bloch}b,d)
	where two TCWs embrace an AB island enabling full control of the quantum state via magnetic and geometric phase manipulations.
}
\he{The potential profile} of both models is shown in Fig.~\ref{fig:TransmSimulations}a,d.
\he{
	For such study of a nanoscale device, materials properties have to be specified in the nextnano input file such that the potential energies are properly set.
	% To study such a nanoscale device, one has to specify proper materials
	% in the nextnano input file. This sets the potential energy height
	% for all regions of the device.
}
% \oy{
% To study such a nanoscale device, one has to specify proper materials
% in the nextnano input file. This sets the potential energy height
% for all regions of the device.
% }
We have used GaAs for the device and leads, and adjusted the potential energy in different regions representing high insulating
barriers (red lines), tunneling barriers (light blue lines), and gates
(green regions).
The potential energy at the gates can be tuned by applying
\he{gate voltages}
% \sout{`gate voltages'} % The "" gives an impression that the software cannot treat true voltages, what is wrong. So better avoid it. I know you want to indicate that it is energy eV and not voltage V.
and the strengths of the barriers are given in Fig.~\ref{fig:TransmSimulations}
and its caption.

% \sout{
% {\it The pLDoS}
% is the probability to find in a given space-point
% the propagating electron
% which has a given energy, $E$, and has been injected e.g. in the first lead.
% We note in passing that the local density of states is the sum of the pLDoS over
% all leads.
% Hence, the coordinate dependence of pLDoS, $ n(x,y, E)$, illustrates how
% the electron with a given energy propagates through the device \cite{datta_2005}.
% %%%
% % \SB{for each lead}.
% %%%
% }
\he{
	Several examples of the pLDoS are shown in Fig.~\ref{fig:density} which have
	been generated using the nextnano software.
	In these simulations, the electron has been injected into lead no.~1 with a given energy that is $ E = 9.2 $~meV for TCW and $ E = 7.5 $~meV for the quantum interferometer (TCW--AB--TCW).
	Slices of the pLDoS are shown at these energies.
}
% \sout{Leads
% $ \lambda = 2 \ldots 4 $,
% with $ \lambda $ being the lead number,
% can also be investigated, if needed.}% H.E.:
% \comment{H.E.: Of course. This information is not important I think. Further, we use lambda nowhere. Rather we should put the discussion of Fig. 12 here. What do we see there? I moved it from the caption:}

	Let us first discuss the TCW (left column of Fig.~\ref{fig:density}) for different voltages ($V_{\rm T}$) applied on the tunnel-barrier gate.
	The TCW is able to change only the angle $ \theta $ -- see Eq.~(\ref{RotState}).
	Thus, the output state can be written as $ | \Psi \rangle_{\rm TCW} \propto
	\cos(\theta / 2) | 0 \rangle + \sin(\theta / 2) | 1 \rangle $.
	When the
	tunneling barrier is absent (Fig.~\ref{fig:TransmSimulations}a), one
	observes at the output an equal superposition of $ | 0 \rangle $ and
	$ | 1 \rangle $ which can occur at either $ \theta = \pi / 2 $ or
	$ \theta = 3 \pi / 2 $, both corresponding to two points on the equator
	of the Bloch sphere (Fig.~\ref{fig:bloch}). Since a small increase of
	$ V_{\rm T} $ drives the output state to $ | 0 \rangle $
	(Fig.~\ref{fig:TransmSimulations}b) corresponding to the north pole
	of the Bloch sphere, we conclude that in barrier-less setup $ \theta = 3 \pi / 2 $.
	% Note also an oscillation of the pLDoS in
	% Fig.~\ref{fig:TransmSimulations}a in the connected wire that is extremely sensitive to the length of the connection. #\cbc{meaning ?}
	With further increasing
	$ V_{\rm T} $, the output state becomes $ | 1 \rangle $, i.e. the south pole of the Bloch sphere (Fig.~\ref{fig:TransmSimulations}c, $ \theta =
	\pi $) and returns to the equator (Fig.~\ref{fig:TransmSimulations}d,
	$ \theta = \pi / 2$). The latter point is the opposite equator point to that of the barrier-less setup.
	When the tunneling barrier becomes high
	(Fig.~\ref{fig:density}d) one enters the regime of two fully decoupled
	transport channels with output state $ | 0 \rangle $. Clearly, these
	coherent tunnel oscillations of the electron wave function manifest itself
	in the quantum oscillations of the transmission via TCW
	(see Fig.~\ref{fig:TransmSimulations}c).
	
	Secondly, let us focus on electron propagation through the quantum interferometer (Fig.~\ref{fig:density}f-j). Increasing \he{the} potential on a side gate of the lower transport channel ($V_{\rm g}$)
	modifies \he{the} geometric phase of the electron's quantum state and
	changes the second angle $ \phi $ in Eq.~(\ref{fig:bloch}).
	This, in turn, causes coherent oscillations between the output terminals
	3 and 4, i.e. oscillations between output states $\ket{0}$ and $\ket{1}$,
	and results in quantum oscillations of the TCW--AB--TCW transmission
	(see Fig.~\ref{fig:TransmSimulations}f).
	The tunnel regions in the example of Figs.~\ref{fig:density}f-j are
	the same and each of them changes the angle
	$ \theta $ by $ \pi / 2 $.
	This is apparent from Fig.~\ref{fig:density}a where $ V_{\rm g} = 0 $ and $ \phi = 0 $: the output state after successive rotation in two TCW regions is $ | 1 \rangle $, i.e. the total change of
	$ \theta $ in the TCWs is $ \pi $. Therefore, each individual connection
	changes $ \theta $ by $ \pi / 2 $. This allows one to approximate the
	output state as $ | \Psi \rangle_{\rm TCW-AB-TCW} \propto (e^{i \phi} - 1)
	| 0 \rangle + (e^{i \phi} + 1) | 1 \rangle $ \cite{bauerle_2018}. Similar to the analysis of the pLDoS in the TCW, we can now trace rotations of
	the electron state with increasing $ V_{\rm g} $. Two output states shown
	in Fig.~\ref{fig:density}g,i correspond to two opposite points on the equator of the Bloch sphere with $ \phi = \pi / 2 $ and $ \phi =
	3 \pi / 2 $. The south pole of the Bloch sphere is reached in Fig.~\ref{fig:density}j despite a substantial blockage of the lower
	transport path by the strong gate potential of the side gate.
	%%%
	% Following the shape in the bottom transport path for the voltage modification % (Fig.~\ref{fig:density}f-j), we observe that the pLDoS is gradually reduced
	% what can potentially block transmission at large side-gate voltage.
	%%%
	
	To conclude the discussion of the pLDoS, we note
that a detailed analysis of the flying qubit geometry in relation with
experiments has also been performed using a combination of the theoretical
approach and the KWANT software, see Ref.~\cite{BautzePRB2014}.

\he{
	Having discussed electron propagation
	at a qualitative level
	via the pLDoS, let us now investigate electron propagation in a more quantitative way employing energy dependent transmission, $T_{ij}(E)$ from the first lead to the output leads no.~3 and no.~4.
	First, we study the dependence on the injection energy  (Fig.~\ref{fig:TransmSimulations}b,e).
}
% \sout{{\it The energy dependent transmission}, $ T_{ij}(E) $,
% is determined by the probability
% for the electron which is injected into lead $ i $ to reach lead $ j $.}
% %%%
% %%% OYe: text moved up
% %%%
% \oy{The simulations of the transmission are shown in Fig.~\ref{fig:TransmSimulations}b,c,e,f.} % H.E.: I think we should put the references step-by-step with proper description ... and not all at once here.
% We have used the following strategy for theses simulations:
% %%%
% % \oy{of transmission}:
% %%%
% First, we have studied transmission from the first lead to the output leads no.~3 and no.~4 for different energies of the injected electron
% (Fig.~\ref{fig:TransmSimulations}b,e).
\he{The energy of the incoming electron is counted from the potential energy of the
	lead no.~1.}
All device parameters are fixed at this stage.
Transmission is zero if the electron energy is smaller than the energy of propagating states of the device,
\he{that is below $ 6.2 \rm{~meV}$}
for the chosen parameters.
\he{Above this threshold},
transmission starts to grow.
% \sout{above this energy threshold.}
However, it is \he{first} accompanied by substantial reflection of the electron to the input leads no.~1 and no.~2
\he{which are described by $ T_{11} \mbox{ and } T_{12} $ (not shown).
	%%%
	% \todo{O.Y.: Enhance how the figures are referred to (a,b,c, ...).
	% clarify energy ranges where total transmission is reduced ...}
	%%%
	This is apparent in the central panels of Fig.~\ref{fig:TransmSimulations}, up to an energy of about \SI{7}{\milli\eV}, where
	the total transmission, $ T_{\rm total} = T_{13} + T_{14} $
	(magenta lines), is smaller than the ideal value, $ T_{\rm total} <
	T_{\rm ideal} = 1 $.}
%%%
% \oy{This is visible in the central panels of
% Fig.~\ref{fig:TransmSimulations}, in the energy range from $ 6.2 \rm{~meV}$
% to $ \sim 7 \rm{~meV}$ where the total transmission is smaller than the ideal
% value, $ T_{13} + T_{14} < 1 $ (magenta lines).}
%%%
\he{In a second step,}
% Next
we have identified the energy at which $ T_{\rm total} \simeq 1 $ (i.e. reflection is minimised) but $ T_{13}, T_{14} \ne 0, 1 $.
% \sout{(expected strong sensitivity of the electron states to the gates and the barriers).}% H.E.: This is important information, right? Let's write it as a proper sentence.
\he{
	In this regime, we expect strong sensitivity of the electron state $ | \psi \rangle $,
	Eq.~(\ref{RotState}), to the gates and the barriers of the quantum device.
}
\he{
	Two examples of these energies are shown by red dashed vertical lines and are studied at the second stage of the simulations.
	Fixing these two energies for the TCW and the quantum interferometer, we have studied the dependence of the transmission on the potential of the tunnel barrier, $V_{\rm T}$ (Fig.~\ref{fig:TransmSimulations}c), and the side gate, $V_{\rm g}$ (Fig.~\ref{fig:TransmSimulations}f).
}
% \oy{
% Two examples of these energies are shown by red dashed vertical lines
% and are studied at the second stage of the simulations. Namely,
% %%%
% % At the second stage of the simulations,
% %%%
% we have focused on these chosen energies and}
%  studied the dependence of the transmission on the tunneling barrier height, Fig.~\ref{fig:TransmSimulations}c, and on the asymmetric gating of the interferometer,
% Fig.~\ref{fig:TransmSimulations}f.
The goal of this stage is
twofold: We identify regions of the parameters where
$ T_{\rm total} $
%%%
% the total transmission
%%%
(magenta lines) is close to the ideal value of 1 and, simultaneously, the rotation of the electron states is pronounced.
The latter condition is fulfilled by the crossover between regimes $ T_{13} < T_{14} $ and
$ T_{13} > T_{14} $.
The vicinity of the crossover can be chosen as an operation range of the qubit (or of the qubit element), provided the reflection is almost absent.
% \sout{The residual reflection can be suppressed by gating out the incoming channels after the electron is injected.} % H.E.: I think this is beyond experimental reality.
\he{
	In order to find optimal parameters which provide both ideal transmission and range for manipulation of the quantum phase, simulations such as the here-discussed case are helpful to identify experimentally relevant voltage ranges.
	The here-discussed simulations indicate that sufficient control is obtained in TCW and AB regions with the length $\geq 500$~nm and further allow to identify optimal operation voltages.
	Let us discuss these aspects in more detail for the TCW and the quantum interferometer.
	% Clearly, to find optimal parameters which provide a combination of the almost ideal
	% transmission and  the feasible manipulation of the quantum phase,
	% %%%
	% % to have a flexibility in choosing optimal gate voltages,
	% %%%
	% one needs pronounced quantum oscillations in the experimentally relevant range of voltages.
	% This can be achieved in relatively long TCW and AB regions
	% (with the length $\geq 500$~nm in the examples which we simulated).
	%%%
	% This requires a noticeable relative phase of electronic modes which govern the quantum
	% contribution to the transmission.
	% Those are two lowest in energy modes in the case of the TCW unit or two modes propagating
	% through the upper and lower arm of the interferometer.
	% In both cases, the relative phase grows with increasing the longitudinal, i.e.
	% in the direction of propagation, size of the nanounit and the functionality of the qubit
	% can be improved by optimizing its length.
	%%%
	% \comment{H.E.: The previous text is not very concisely written. It is hard to understand.}
	%%%
	% Let us explain this for both setups.
	%%%
	
	{\it TCW}: A simple phenomenological scattering theory predicts that
	$ T_{13} \simeq \cos^2( \delta \phi_{\rm TCW} / 2 ) $, $ T_{14} \simeq
	\sin^2( \delta \phi_{\rm TCW} / 2 ) $,
	$ \delta \phi_{\rm TCW} \equiv \delta k L_{\rm TCW} $, where $ \delta k = k_1 - k_2 $ is the
	difference of wave vectors of two modes with the lowest energy, which
	support the transmission, and $ L_{\rm TCW} $ is the effective spatial scale
	of the region, where the tunneling takes place \cite{bauerle_2018}.
	The quantum phase $ \delta \phi_{\rm TCW} $ is expected to grow with
	increasing the length of the tunneling barrier, $ L_{\rm tun} $, which is equal
	to 300 nm in the example of the left column of Fig.~\ref{fig:TransmSimulations}.
	We distinguish $ L_{\rm TCW} $ and $ L_{\rm tun} $ since the former depends
	on the shape of the electrostatic potential inside the device.
	%%%
	% and on the pLDoS inside the TCW region.
	%%%
	Therefore, one may expect
	$ L_{\rm TCW} \ll L_{\rm tun} $. This inequality has been confirmed by a
	comparison of the scattering theory (green and cyan dots
	in Fig.~\ref{fig:TransmSimulations}c) with the outcome of the true 2D simulations
	(solid lines in the same figure). $ L_{\rm TCW} $ is the only adjustable parameter
	of this comparison. The value of $ \delta k $ has been found by using
	the dispersion relation of almost free electrons propagating in a
	semiconductor, $ E_{1,2} = (\hbar k_{1,2})^2 / 2 m^* $. Energy levels
	$ E_{1,2} $ have been obtained from
	1D simulations of the spectrum at the 1D transverse cross section in the center
	of the device (dashed line in Fig.~\ref{fig:TransmSimulations}a).
	%%%
	% while two lowest energy levels have been found from.
	%%%
	% We have found an excellent agreement between
	% scattering theory and the 2D simulations at $ L_{\rm TCW} = \ldots L_{\rm tun}
	% \ll L_{\rm tun} $. \comment{H.E.: Conversion factor missing.}
	%%%
	Interestingly, the ratio $ L_{\rm TCW} / L_{\rm tun} $ is almost insensitive to
	the transverse size of the device. An excellent agreement between the 2D simulations
	and the scattering theory suggests that the latter can be used as a compact model
	of TCW in simulations of more complicated circuits. Such a simplification will allow
	one to minimize computer resources needed for the simulations. The inset in Fig.~\ref{fig:TransmSimulations}c
	shows that the range of $ V_{\rm T} $,
	where the quantum oscillations occur, shrinks with making $ L_{\rm tun}$, and
	correspondingly the space for the quantum interference, smaller. Since
	$ T_{\rm total} $ is very close to 1 (no reflection) in the entire range $ 0 < V_{\rm T}
	< 1~\rm{eV}$, such an idealized qubit would operate properly in a vicinity of any crossover point where $ T_{13} \simeq T_{14} $, e.g. $ V_{\rm T} \sim 0.1~\rm{eV}$
	or $ V_{\rm T} \sim 0.33~\rm{eV}$.
	
	{\it Interferometer}: When the electron modes propagate through the upper
	and lower arms of the electrostatic version of the Aharanov--Bohm interferometer,
	they acquire a relative phase which governs the quantum interference.
	If the interferometer is connected directly to the leads, the transmission through
	the device can be estimated as $ {\cal T}_{\rm AB} = \cos^2 \left( e \, \delta V
	\tau_{\rm AB} / 2 \hbar \right) $ \cite{datta_1986}. Here, $ \tau_{\rm AB} $ is the
	flight time of the electron through the unit. In the ballistic case, it
	is the ratio of the interferometer length over the electron velocity,
	$ \tau_{\rm AB} = L_{\rm AB}/v$. We have introduced the relative total
	potential, $\delta V = V_{\rm{u}} - V_{\rm{l}} $
	(integrated over the upper, $V_{\rm{u}}$, or lower, $V_{\rm{l}}$, arm), which
	the electron feels inside the interferometer.
	The interference oscillations are more pronounced in longer devices,
	%%%
	% \comment{H.E.: It is the Aharanov--Bohm effect.}
	%%%
	% This is illustrated by the right lower panel of Fig.~\ref{fig:TransmSimulations},
	%%%
	compare results presented in Fig.~\ref{fig:TransmSimulations}f and its inset.
	%%%
	% The simulations were done for the interferometer connected to two TCW regions.
	%%%
	To avoid complexity, we do not discuss
	here semi-phenomenological analytical calculation of transmission for the composite
	TCW--AB--TCW device and do not compare the scattering theory with the 2D simulations. Similar
	to the TCW device, it is useful to tune $ V_{\rm g} $ to the vicinity of the
	crossover point where $ T_{13} \simeq T_{14} $, i.e. either $ V_{\rm g} \sim 0.1~{\rm eV}$
	or $ V_{\rm g} \sim 0.5~{\rm eV}$ in the example of Fig.~\ref{fig:TransmSimulations}f.
	Note, however, that $ T_{\rm total} < 1 $ in both cases.
	%%%
	% there is a reflection.
	%%%
	Clearly, the preference should be given to the regime with smaller reflection, i.e. the
	second crossover point in the simulated example.
	%%%
	% We note that undesirable decoherence can become pronounced in the long
	% devices. Since decoherence spoils the quantum information and the interference,
	% the choice of the proper length is really the optimization between constructive
	% and destructive effects.
	%%%
}

%%%
% {\color{red}TO BE UPDATED BY OYe: Possible options for the proper operation range
% are $ 0.15\rm{~eV} < V_T < 0.25\rm{~eV} $ for the first device and $ 0.2\rm~{eV} <
% $ V_g < 0.6\rm{~eV} $ for the second one.
% }
%%%

% \oy{
% %%%
% % I think here we have been discussing Fig. 13 enough. After this paragraph
% % I would only put a short discussion and conclusion section.
% %%%
% The above predictions of numerical simulations can serve as an important input for experimental realisations of the flying qubit and can be used in a feedback loop of the workflow (Fig.~\ref{fig:nn_workflow}) to find the perfect geometry for the devices tailored to the different approaches, such as Levitons and SAW. % Avoiding the word "can" makes the content much more credible ...
% Though details might vary in the different approaches, finding the optimal parameters of the generic theoretical models is valuable for engineering real devices.}
\he{
	The above predictions of numerical simulations serve as important input for experimental realisations of the flying qubit.
	Integrating the simulation additionally in a feedback loop of the workflow (Fig.~\ref{fig:nn_workflow}) would enable to find optimised device geometries tailored to the different approaches such as Levitons or SAW.
}
To conclude this section, we would like to mention that the applications of the nextnano
software can be very broad since it can straightforwardly be adapted for modelling devices made from other semiconductor materials (e.g., industrially highly relevant
SiGe), for including effects of the magnetic field on the interferometer, for mimicking dephasing and decoherence with the help of the artificially connected leads, to name just a few.

\subsection{Low energy time-resolved and many-body simulations}

In the previous section, we have focused on the simulation of the static properties of
the devices. Solving the corresponding quantum-electrostatic problem allows one to understand
how macroscopic parameters, usually the geometry of the electrostatic gates that are set to typical
values of order $ \leq 1{\rm ~eV}$,
influence the active quantum part of the device where the relevant energies are in the $\rm meV$ range.
Once this is understood, the next step is to simulate actual time-resolved experiments that involve
sub-meV physics (typical time in the $1-100 {\rm ~ps}$ range). The goal is to understand the propagation
of pulses, the coupling between different pulses (at the origin of the two-flying-qubits gates),
the renormalisation of the velocity due to Coulomb interactions \cite{roussely_2018} and other effects such as
different decoherence and relaxation channels. These theoretical developments are very
much on-going research for which no standard approaches have yet emerged.
\he{As}
in-depth discussion of
these aspects goes beyond the \he{scope of the} present review, we refer to Ref.~\cite{bauerle_2018}
for pointers to the
literature or to Ref.~\cite{gaury_2014} for an introduction to the non-interacting formalism and to Ref.~\cite{kloss_2021} for illustrative time-dependent simulations of the propagation of voltage pulses,
in particular in the quantum Hall regime \cite{gaury_2014b}.

These methods have not been included into the nextnano software, however, TKWANT, the time-dependent extension
of the KWANT software is able to provide an appropriate platform which is complementary to the nextnano one.
Both KWANT and TKWANT software packages are distributed under the BSD license which imposes minimal restrictions
on the use and distribution of this software. Consequently, algorithms developed by the KWANT team could be
incorporated into commercial software packages targeting specific quantum industry applications, such as electron-flying-qubit devices.
The principal developers of KWANT are from CEA Grenoble and TU Delft.

Note that there are no general purpose simulation approaches that can handle many-body problems in an exact and systematic way except in very particular cases. Most approaches rely on some approximation scheme whose validity must be checked a posteriori. A promising route followed by some of us to design a systematic method with a controlled accuracy uses calculations of high order processes (i.e. processes where electrons interact strongly) made possible by the use of a machine learning approach to evaluate the corresponding
high dimensional integrals \cite{macek_2020}.

\section{Conclusion and outlook}

The realisation of flying qubits with single electrons opens a novel, viable route of quantum technology with considerable potential for quantum-computation applications.
In this review we introduced the novel electron-flying-qubit approach and discussed three equally promising transport techniques -- surface acoustic waves (SAW), hot-electron emission from quantum-dot pumps and Levitons -- which are rapidly advancing.
\he{Owing to similarities between the different approaches -- such as emission from a gate-defined quantum dot in SAW-transport and the electron pump -- we suspect that progress in one-field will also drive the others.}
Based on latest progress and relevant simulation cases, we showed that numerical modelling of quantum devices is decisive to speed up experimental deployment cycles towards the first implementation of an electron flying qubit.
We anticipate that automatised optimisation of the device design via numerical modelling will enable nanofabrication tailored for efficient quantum operations.

In order to make the electron flying qubit competitive with cutting-edge approaches in the field of quantum computation, it is of central importance to develop ultrafast real-time control of quantum operations.
An appealing approach to implement such in-flight quantum operations is to use ultrafast voltage pulses in the picosecond range and below.
On-chip optoelectronic conversion of a femtosecond laser
pulse is so far the most promising technique to generate electrical pulses on the picosecond scale \cite{Auston1975, Mourou1981, Auston1984}.
Combined with recent conversion efficiency improvements of these optoelectronic devices \cite{heshmat2012, Lepeshov2017,Bashirpour2017, giorgos_2020}, such a real-time control is in reach and is currently pursued in the {\it UltraFastNano} project.
Using these techniques, single-electron wave packets with a temporal width of 1 ps can be generated.
The thus-enabled miniaturisation of quantum interferometers will allow the implementation of hundreds of quantum operations within the coherence time.
Furthermore, ultrafast gate control will provide a possibility to resolve quantum states in real time.
Rather than measuring the coherent oscillations of the electron qubits by varying the strength of the tunnel coupling \cite{yamamoto_2012}, one can simply control the tunnel barrier in a time-resolved manner.
This enables to keep the electrostatic confinement potential of the entire device constant and only vary the tunnel barrier on the time scale needed for the quantum operation.

The progress in the field
% \sout{also strongly}
\he{strongly}
depends on the availability of tools for the reliable modelling of
the quantum devices.
The simulations must possess
\he{enough predictive power to suggest}
% \sout{ a predictive power which could suggest}
the most suitable device geometry prior to the fabrication of the device in a clean room.
Iterations and tests of the devices are costly and time consuming and should be reduced to
the strict minimum with the help of the high-precision professional simulations.
Adding more and more qubits into quantum circuits will increase drastically the experimental
parameter space for device tuning. Therefore, automatic tuning of all the gate voltages by using
concepts from artificial intelligence and machine-learning would have to be implemented in platforms for the theoretical modelling.
\he{We anticipate that} the synergy of semiconductor quantum technology with cutting-edge numerical simulations paves
% \sout{thus}
the way for electron-flying-qubit implementations fostering the industrial applicability of quantum computation.
	
\section*{Acknowledgements}
C.B. acknowledges helpful discussions with J.-M. G\'erard.

\section*{Funding}
We acknowledge funding from the European Union's Horizon 2020 research and innovation programme under grant agreement No. 862683 (UltraFastNano).
E.C. acknowledges funding from the European Union’s Horizon 2020 research and innovation programme under the Marie Sk\l{}odowska-Curie grant agreement No. 840550 (PRESQUE).
J.W. acknowledges funding from the European Union’s Horizon 2020 research and innovation programme under the Marie Sk\l{}odowska-Curie grant agreement No. 754303 (GreQuE). C.B., D.C.G. and X.W. acknowledge funding from the French National Funding Agency (ANR) through project ANR-16-CE30-0015-02 (FullyQuantum). C.B. and G.G. acknowledge funding from the French National Funding Agency (ANR) through  project ANR- 19-CE47-0005 (STEPforQUBITS).

\section*{Availability of data and materials}
Numerical codes and tutorials can be found at https://kwant-project.org and https://www.nextnano.com.

%%%
% \section*{Competing interests}
% The authors declare that they have no competing interests.
%
% \section*{Author contributions}
% H.E., J.W., E.C., M.C.d.S.F., M.K., J.S., X.W., D.C.G., S.B., A.T., T.G., O.Ye. and C.B. wrote the article with input from all authors.
% T.C., G.G. and S.O. contributed to the device fabrication. P.P. contributed to the experimental set-up.
% H.E., J.W., E.C., M.C.d.S.F, C.G., A.L., P.R., K.\"O., M.A. prepared the figures.
%%%

\bibliographystyle{bmc-mathphys}
\bibliography{EPJ_QT-QI}        % Bibliography file (usually '*.bib' )

\newpage

\section*{Figures}

% \begin{figure}[!h]
%  \centering
%  \includegraphics[width=11cm]{fig10_calibration.png}
% 	\caption{
% 	\textsf{\textbf{Calibration details}
%      (a) Schematic design of the heterostructure (left) below the QPC, showing also the location of the surface charges and the 2DEG, QPC design (middle) from scanning electron microscopy pictures, and its location on the chip (right). (d) Experimental results of the current response to a symmetric bias, V\textsubscript{G}, on both gates of the QPC shown in (a). (b, c, e) Electron density in the 2DEG, in the region below the QPC gates (white dots) at different bias conditions. The gates are superimposed on the density, but they are located on the surface.}
% 	}
% 	\label{simulations}
% \end{figure}

\end{document}